\documentclass[a4paper,12pt]{article}
\pdfoutput=1
\usepackage[margin=1in]{geometry}
\usepackage{graphicx,xcolor}
\usepackage{mathtools,braket,blkarray}
\allowdisplaybreaks
\usepackage{amssymb,amsfonts,bbm,relsize}
\usepackage[width=0.9\textwidth,footnotesize]{caption}
\usepackage[font=scriptsize]{subcaption}
\usepackage{cite,hyperref,url}
\hypersetup{pageanchor=false}
\usepackage{booktabs,multirow,makecell}
\usepackage[utf8]{inputenc}
\usepackage{cleveref}
\usepackage{amsmath}

\begin{document}

\begin{titlepage}
\begin{center}
{\bf\Large Multiple modular symmetries as the origin of flavour
  } \\[12mm]
Ivo~de~Medeiros~Varzielas$^{\dagger}$%
\footnote{E-mail: \texttt{ivo.de@udo.edu}}, 
Stephen~F.~King$^{\star}$%
\footnote{E-mail: \texttt{king@soton.ac.uk}}, 
Ye-Ling~Zhou$^{\star}$%
\footnote{E-mail: \texttt{ye-ling.zhou@soton.ac.uk}}
\\[-2mm]

\end{center}
\vspace*{0.50cm}
\centerline{$^{\dagger}$ \it
CFTP, Departamento de F\'{i}sica, Instituto Superior T\'{e}cnico, Universidade de Lisboa,}
\centerline{ \it
 Avenida Rovisco Pais 1, 1049 Lisboa, Portugal}
 \vspace*{0.2cm}
\centerline{$^{\star}$ \it
School of Physics and Astronomy, University of Southampton,}
\centerline{\it
SO17 1BJ Southampton, United Kingdom }
\vspace*{1.20cm}

\begin{abstract}
{\noindent
We develop a general formalism for multiple moduli and their associated modular symmetries.
We apply this formalism to an example based on 
three moduli with finite modular symmetries $S_4^A$, $S_4^B$
and $S_4^C$, associated with two right-handed neutrinos and the charged lepton sector, respectively.
The symmetry is 
broken by two bi-triplet scalars to the diagonal $S_4$ subgroup.
The low energy effective theory involves the 
three independent moduli fields $\tau_A$, $\tau_B$ and $\tau_C$,
which preserve the residual modular subgroups $Z_3^A$, $Z_2^B$ and $Z_3^C$, in their respective sectors,
leading to trimaximal TM$_1$ lepton mixing, consistent with current data, without flavons.
}
\end{abstract}
\end{titlepage}

\section{Introduction}

The discovery of neutrino mass and mixing implies that the Standard Model (SM) must be extended somehow.
An elegant possibility remains 
the original type Ia seesaw mechanism~\cite{Minkowski:1977sc, Yanagida:1979ss, Gell-Mann:1979ss, Glashow:1979ss, Mohapatra:1979ia,Schechter:1980gr, Schechter:1981cv} involving right-handed neutrinos, which, when integrated out, 
yield the Weinberg operators $HHL_iL_j$, where $H$ is the Higgs doublet of the SM and $L_i$
is a lepton doublet of the $i$th family. 
The minimal type Ia seesaw mechanism supplements the particle content of the SM by 
just two right-handed neutrinos (2RHN)~\cite{King:1999mb,King:2002nf},
and this approach will be followed in the present paper.
However, to explain the observed approximate tri-bimaximal 
lepton mixing, one must go beyond the seesaw mechanism and 
consider a non-Abelian discrete family symmetry~\cite{King:2013eh,King:2017guk}.
For example, $S_4$ has been used to account for trimaximal TM$_1$ lepton mixing
\cite{Varzielas:2012pa, Luhn:2013vna}, enforced by a residual $Z^{SU}_2$ symmetry in the neutrino sector, and a residual
$Z_3^T$ in the charged lepton sector~\footnote{We adopt the standard presentation of the 
$S_4$ generators $S,T,U$
where $S^2=T^3=U^2=(ST)^3=(SU)^2=(TU)^2=(STU)^4=I$~\cite{King:2013eh}.}.
However such realistic models typically involve many flavons.

The origin of such non-Abelian discrete family symmetry might be 
due to a continuous non-Abelian gauge symmetry
\cite{deMedeirosVarzielas:2005qg, Koide:2007sr, Banks:2010zn, Luhn:2011ip, Merle:2011vy, Wu:2012ria, Rachlin:2017rvm, King:2018fke}.
Alternatively, it could be due to extra dimensions \cite{Asaka:2001eh, Altarelli:2006kg, Kobayashi:2006wq, Altarelli:2008bg, Adulpravitchai:2009id, Burrows:2009pi, Adulpravitchai:2010na, Burrows:2010wz, deAnda:2018oik, Kobayashi:2018rad, deAnda:2018yfp, Baur:2019kwi}. 
With extra dimensions, it could either arise as an accidental symmetry
of the orbifold fixed points
(for recent discussion with two extra dimensions, see  \cite{Kobayashi:2008ih,deAnda:2018oik,Olguin-Trejo:2018wpw, Mutter:2018sra}) or as a subgroup of the symmetry of the extra dimensional lattice, known as modular symmetry~\cite{Giveon:1988tt}, arising from superstring theory \cite{Ferrara:1989bc,Ferrara:1989qb}
\footnote{The geometric connection between the origin of the family symmetry due to 
modular symmetry and the orbifolding method with two extra dimensions
has recently been discussed, e.g., in \cite{deAnda:2018ecu, Kobayashi:2018bff}. On the other hand, massive states predicted in string theories may break the modular symmetries. This effect is naturally suppressed by the Planck scale, and thus can be safely ignored. }.
Indeed, it has been suggested that a finite subgroup of the modular symmetry group, when interpreted as a family symmetry, might help to provide a possible explanation for the neutrino mass matrices
\cite{Altarelli:2005yx, deAdelhartToorop:2011re}, and this will be the approach followed here.

Recently it has been suggested that finite modular symmetry might be the origin of flavour mixing with 
neutrino masses as modular forms
\cite{Feruglio:2017spp}, leading to constraints on the Yukawa couplings.
This has led to a revival of the idea that  modular symmetries are symmetries of the extra dimensional spacetime
with Yukawa couplings determined by their modular weights \cite{Criado:2018thu}.
The finite modular groups $\Gamma_2\simeq S_3$~\cite{Kobayashi:2018vbk,Kobayashi:2018wkl}, 
$\Gamma_3\simeq A_4$~\cite{Feruglio:2017spp,Criado:2018thu,Kobayashi:2018scp,Okada:2018yrn,Kobayashi:2018wkl,Novichkov:2018yse}, $\Gamma_4\simeq S_4$~\cite{Penedo:2018nmg,Novichkov:2018ovf} and 
$\Gamma_5\simeq A_5$~\cite{Novichkov:2018nkm,Ding:2019xna}
have been considered, in which special Yukawa structures are consequences of the modular forms. Compared with
traditional neutrino models of flavour symmetry, only a minimal set of flavon fields (or no flavons at all) 
need to be introduced in the new framework~\footnote{Extension to the quark flavour mixing is given in \cite{Kobayashi:2018wkl,Okada:2018yrn, Okada:2019uoy}.}, making such an approach very attractive.

Within the framework of finite modular symmetry outlined above, only a single modulus field $\tau$ is usually considered,
corresponding to a single finite modular symmetry $\Gamma_N$.
It has been pointed out that particular modular forms, corresponding to special values of $\tau$,
preserve a residual subgroup of the finite modular symmetry $\Gamma_N$. 
For example, such residual symmetries are considered in  \cite{Novichkov:2018yse} as subgroups of the modular $A_4$ symmetry. 
Some of these specific values for $\tau$ have been shown to be obtained in extra dimensions through orbifolding \cite{deAnda:2018ecu}.
With the help of two moduli with different 
residual symmetry $Z_3$ in the charged lepton sector and $Z_2$ in the neutrino sector, it was shown how 
trimaximal TM$_2$ lepton mixing may be realised~\cite{Novichkov:2018yse}. Also brief discussion on residual symmetry after modular $S_4$ symmetry breaking is given in \cite{Novichkov:2018ovf}. 
However, the formalism for having two or more moduli fields (as necessary for such a scheme) has not so far been developed, providing one of the main motivations for the present paper.

In the present paper, we shall extend the formalism of finite modular symmetry to the case of multiple
moduli fields $\tau_J$ ($J=1, \ldots M$) associated with the finite modular symmetry
$\Gamma_{N_1}^1\times \Gamma_{N_2}^2 \times \cdots \times \Gamma_{N_M}^M$.
As an example, we shall then present the first consistent 
example of a flavour model of leptons with multiple modular $S_4$ symmetries
interpreted as a family symmetry.
The considered model involves three finite modular symmetries $S_4^A$, $S_4^B$
and $S_4^C$, associated with two right-handed neutrinos and the charged lepton sector, respectively,
broken by two bi-triplet scalars to their diagonal subgroup. The low energy effective theory consists
of a single $S_4$ modular symmetry with three independent modular fields $\tau_A$, $\tau_B$ and $\tau_C$,
which preserve the residual modular subgroups $Z_3^A$, $Z_2^B$ and $Z_3^C$, 
in their respective sectors~\footnote{Having a separate residual symmetry associated with each of the two right-handed neutrinos and the charged lepton sector was also assumed in  
the tridirect CP approach~\cite{Ding:2018fyz,Ding:2018tuj}, although here we do not assume any 
(generalised) CP symmetry. An extension of modular symmetry 
to include general CP symmetries was given in \cite{Novichkov:2019sqv}.},
leading to trimaximal TM$_1$ lepton mixing, consistent with current data, without requiring any flavons.

The remainder of the paper is organised as follows.
In section~\ref{multiple} we show how the formalism of finite modular symmetry with a single modulus field can be extended to include multiple moduli and an extended finite modular group.
In section~\ref{S4} we have focussed on the case of the single finite modular $S_4$ symmetry,
and have analysed its stabilisers and resulting remnant symmetries. 
In section~\ref{model} we have proposed a model based on three moduli fields associated with a high energy 
finite modular group $S_4^3$, which is broken to a single diagonal $S_4$ with three independent moduli fields at low energies, whose stabilisers leads to different remnant symmetry in the different sectors, which may be used to enforce 
trimaximal TM$_1$ mixing, leading to good numerical fits to the data, once right-handed neutrino mixing is taken into account. Section~\ref{conclusion} concludes the paper.

%%%%%%%%%%%%%%%%%

\section{From single to multiple modular symmetries}
\label{multiple}

Modular invariant supersymmetric field theories have been analyzed in \cite{Ferrara:1989bc,Ferrara:1989qb}. 
Modular invariance is involved in string compactifications and realistic Yukawa couplings arise from modular forms \cite{Ibanez:1986ka,Casas:1991ac,Lebedev:2001qg,Kobayashi:2003vi}. 
It has been invoked while addressing several aspects of the flavour problem in model building \cite{Brax:1994kv,Binetruy:1995nt,Dudas:1995eq,Dudas:1996aa,Leontaris:1997vw,Dent:2001cc,Dent:2001mn}. 
Direct application of modular symmetry to explain lepton flavour mixing was suggested in 
\cite{Feruglio:2017spp}. 
In the rest of this section, we will give a short review of effective modular-invariant supersymmetry and then expand the formulism to include multiple moduli fields. 

\subsection{A single modular symmetry }

The modular group $\overline{\Gamma}$ acting on the complex modulus $\tau$ (${\rm Im}(\tau)>0$) as linear fractional transformations:
\begin{eqnarray} \label{eq:modular_transformation}
\gamma: \tau \to \gamma \tau = \frac{a \tau + b}{c \tau + d}\,,
\end{eqnarray}
where $a, b, c, d$ are integers and satisfy $ad-bc=1$. 
It is convenient to represent each element of $\overline{\Gamma}$ by a two by two matrix \footnote{Note that it need not be a unitary matrix.}. Then, $\overline{\Gamma}$ is expressed as 
\begin{eqnarray}
\overline{\Gamma} = \left\{ \begin{pmatrix} a & b \\ c & d \end{pmatrix} / (\pm \mathbf{1})\,,~ a, b, c, d \in \mathbb{Z}, ~~  ad-bc=1  \right\} \,.
\end{eqnarray}
This group is isomorphic to the projective special linear group $PSL(2,\mathbb{Z}) = SL(2,\mathbb{Z})/\mathbb{Z}_2$. 
The modular group has two generators, $S_\tau$ and $T_\tau$, which satisfy $S_\tau^2 = (S_\tau T_\tau)^3 = \mathbf{1}$. They act on the modulus $\tau$ and take the following forms
\begin{eqnarray}
S_\tau: \tau \to -\frac{1}{\tau} \,, \hspace{1cm}
T_\tau: \tau \to \tau + 1\,,
\end{eqnarray}
respectively. Representing them by two by two matrices, we obtain 
\begin{eqnarray}
S_\tau=\begin{pmatrix} 0 & 1 \\ -1 & 0 \end{pmatrix}\,, \hspace{1cm}
T_\tau=\begin{pmatrix} 1 & 1 \\ 0 & 1 \end{pmatrix} \,.
\end{eqnarray}

$\overline{\Gamma}$ is a discrete but infinite group. By requiring $a, d = 1~({\rm mod}~N)$ and $b, c =  0~({\rm mod}~N)$, $N=2, 3, 4, \cdots$, i.e., 
\begin{eqnarray} \label{eq:mode_N}
a = k_a N+1\,,~ 
d = k_d N +1\,,~
b = k_b N\,, ~~~~
c = k_c N\,,
\end{eqnarray}
where $k_a$, $k_b$, $k_c$ and $k_d$ are integers, we obtain a subset of $\overline{\Gamma}$ which is also an infinite group and is labelled as 
\begin{eqnarray}
\overline{\Gamma}(N) = \left\{ \begin{pmatrix} a & b \\ c & d \end{pmatrix} \in PSL(2,\mathbb{Z}), ~~  \begin{pmatrix} a & b \\ c & d \end{pmatrix} = \begin{pmatrix} 1 & 0 \\ 0 & 1 \end{pmatrix} ~~ ({\rm mod}~ N) \right\} \,.
\end{eqnarray}
The quotient group $\overline{\Gamma}/\overline{\Gamma}(N)$, labelled as $\Gamma_N$, is a finite group,
also called the finite modular group. 
The finite modular group $\Gamma_N$ can be also obtained by imposing an additional condition for $T_\tau$, $T_\tau^N = \mathbf{1}$, which can be achieved to identify $\tau=\tau+N$ in the upper complex plane \footnote{Note that once $\tau=\tau+N$ is imposed, $\tau' = \frac{-1}{\tau} =  \frac{-1}{\tau+4} = \frac{-\tau}{4\tau -1}$ is automatically satisfied. }.
For $N$ taking some small number, $\Gamma_N$ is isomorphic to a permutation group, 
in particular, $\Gamma_2 \simeq S_3$, $\Gamma_3 \simeq A_4$, $\Gamma_4 \simeq S_4$ and $\Gamma_5 \simeq A_5$ \cite{deAdelhartToorop:2011re}. 

In a theory satisfying the $\Gamma_N$ modular symmetry, any
chiral superfield  $\phi_i$, as a function of $\tau$ (but does not need to be modular forms), non-linearly transforms as \cite{Ferrara:1989bc},
\begin{eqnarray}
 \phi_i(\tau) \to \phi_i(\gamma\tau) = (c\tau + d)^{-2k_i} \rho_{I_i}(\gamma) \phi_i(\tau)\,,
 \label{eq:field_transformation}
\end{eqnarray}
where $-2k_i$ with $k_i$ an integer is the modular weight of $\phi_i$, ${I_i}$ is the representation of $\phi_i$ and $\rho_{I_i}(\gamma)$ denotes a unitary representation matrix of $\gamma$ with $\gamma$ an element of $\Gamma_N$.

Considering an $\mathcal{N}=1$ supersymmetric model in the finite modular symmetry, the action in general takes the form \cite{Ferrara:1989bc, Ferrara:1989qb}
\begin{eqnarray}
\mathcal{S} = \int d^4x d^2\theta d^2\overline{\theta} K(\phi_i, \overline{\phi}_i;\tau,\overline{\tau}) + \left[ \int d^4x d^2\theta W(\phi_i;\tau)+ {\rm h.c.} \right]\,,
\end{eqnarray}
where $h$ is a positive constant. The K\"ahler potential $K$ can be changed at most by a K\"ahler transformation under $\Gamma_N$, and the superpotential $W$ is required to be invariant, i.e.,
\begin{eqnarray}
K(\phi_i, \overline{\phi}_i;\tau,\overline{\tau}) &\to& K(\phi_i, \overline{\phi}_i;\tau,\overline{\tau}) + f(\phi_i, \tau) + \overline{f}(\overline{\phi}_i, \overline{\tau}) \,,\nonumber\\
W(\phi_i;\tau) &\to& W(\phi_i;\tau) \,. 
\end{eqnarray} 

An example of the K\"ahler potential satisfying the K\"ahler transformation takes the following form \footnote{The effects of taking a different form for the Kahler potential are expected to be subdominant, analogously to the results shown by studies of Kahler corrections e.g. \cite{King:2004tx}. Corrections to the Kahler potential may further lead to the stabilisation of the moduli vacua (see, e.g., reviews \cite{Balasubramanian:2004uy,Silverstein:2004id}). We, following all other papers on modular symmetries, avoid this problem by fixing moduli VEVs at typical values.}, 
\begin{eqnarray}
K(\phi_i, \overline{\phi}_i;\tau,\overline{\tau}) = - h \log(-i\tau + i \overline{\tau}) + \sum_{i}  \frac{\overline{\phi}_i \phi_i}{(-i \tau + i \overline{\tau})^{2k_i}} 
\,.
\end{eqnarray}
After $\tau$ gets a vacuum expectation value
(VEV), the K\"ahler potential leaves kinetic terms for the scalar components of the supermultiplets $\phi_i$ and the modulus field as \footnote{The scalar component of $\phi_i$ may gain a non-zero VEV, and this VEV also contributes to the kinetic term of $\tau$. We ignore such a contribution by assuming $v_{\phi_i} \ll \sqrt{h}$. }
\begin{eqnarray}
\frac{h}{\langle -i\tau + i \overline{\tau} \rangle^2} \partial_\mu \overline{\tau} \partial^\mu \tau + 
\sum_i \frac{\partial_\mu \overline{\phi}_i \partial^\mu \phi_i}{\langle -i\tau + i \overline{\tau} \rangle^{2k_i}}  \,. 
\end{eqnarray}

The superpotential $W(\phi_i;\tau)$ is in general a function of the modulus $\tau$ and superfieds $\phi_i$. Under the modular transformation, the superpotential should be invariant under the modular transformation \cite{Ferrara:1989bc}. 
Expanding the superpotential $W(\phi_i;\tau)$ in powers of $\phi_i$, we obtain
\begin{eqnarray}
W(\phi_i;\tau) = \sum_n \sum_{\{i_1, \cdots, i_n\}} \sum_{I_Y} \left( Y_{I_Y} \phi_{i_1} \cdots \phi_{i_n} \right)_{\mathbf{1}} \,.
\end{eqnarray}
Here, $Y_{I_Y}$ represents a collection of coefficients of the relevant couplings. It transforms as a multiplet modular form of weight $2k_Y$ and representation $I_Y$, 
\begin{eqnarray} \label{eq:form_transformation}
Y_{I_Y}(\tau) \to Y_{I_Y}(\gamma \tau) = (c\tau + d)^{2k_Y} \rho_{I_Y}(\gamma) Y_{I_Y}(\tau) \,,
\end{eqnarray}
where $k_Y = k_{i_1} + \cdots + k_{i_n}$ is required to be a non-negative integral. Its representation and weight are required for the invariance of the operator under the $\Gamma_N$ modular transformation. 

\subsection{Multiple modular symmetries \label{sec:multi_modular}}

All lepton flavour models based on finite modular symmetries in the literature so far have been limited to the case of a single modulus field. 
No theoretical approach or model has so far managed to include more than one modulus fields in a self-consistent approach, although the latter case has been briefly mentioned in some references, e.g.~\cite{Novichkov:2018ovf}. In this subsection, we will discuss how to include multiple moduli fields consistently. 

We start from the modular transformation as a series of modular groups $\overline{\Gamma}^{1}$, $\overline{\Gamma}^{2}$, ..., $\overline{\Gamma}^{M}$, where the modulus field for each modular symmetry $\overline{\Gamma}^{J}$ for $J=1,..., M$ is denoted as $\tau_J$. Following Eq.~\eqref{eq:modular_transformation}, any modular transformation $\gamma_J$ in $\overline{\Gamma}^J$ takes the form as 
\begin{eqnarray} 
&&\gamma_J: \tau_J \to \gamma_J \tau_J = \frac{a_J \tau_J + b_J}{c_J \tau_J + d_J} \,.
\end{eqnarray}
A series of finite modular groups $\Gamma_{N_J}^J$ for $J = 1,2,...,M$ can be obtained by modding out an integer $N_J$ by following the discussion in the former section. Note that $N_J$ does not need to be identical to $N_{J'}$ for $J\neq J'$. 

For any finite modular transformations ${\gamma_1, ..., \gamma_M}$ in $\Gamma_{N_1}^1\times \Gamma_{N_2}^2 \times \cdots \times \Gamma_{N_M}^M$, the chiral superfield $\phi_i$, as a function of $\tau_1$, ..., $\tau_M$, now transforms as 
\begin{eqnarray}
 \phi_i(\tau_1, ...,\tau_M) &\to& \phi_i(\gamma_1\tau_1, ..., \gamma_M \tau_M) \nonumber\\
 &&= \prod_{J=1,...,M} (c_J\tau_J + d_J)^{-2k_{i,J}} \bigotimes_{J=1,...,M} \rho_{I_{i,J}}(\gamma_J) \phi_i(\tau_1, \tau_2, ...,\tau_M)\,,
 \label{eq:field_transformation2}
\end{eqnarray}
where $k_{i,J}$ and $I_{i,J}$ are the weight and representation of $\phi_i$ in $\Gamma_{N_J}^{J}$, respectively, and  $\bigotimes$ represents the outer product of the representation matrices for $\rho_{I_{i,1}}$, $\rho_{I_{i,2}}$, ..., $\rho_{I_{i,M}}$. 

For an $\mathcal{N}=1$ supersymmetric model in a series of modular symmetries, the action is extended to the form 
\begin{eqnarray}
\mathcal{S} = \int d^4x d^2\theta d^2\overline{\theta} K(\phi_i, \overline{\phi}_i;\tau_1,...,\tau_M,\overline{\tau}_1,...,\overline{\tau}_M) + \int d^4x d^2\theta W(\phi_i;\tau_1,..., \tau_M)+ {\rm h.c.}\,,
\end{eqnarray}
where $h$ is a positive constant. The superpotential $W$ is required to be invariant under all modular transformations and that the K\"ahler potential $K$ can be changed at most by K\"ahler transformations. 

Including multiple modulus fields, the  K\"ahler potential can be written as, 
\begin{eqnarray}
\hspace{-5mm}
K(\phi_i, \overline{\phi}_i;\tau_1,...,\tau_M,\overline{\tau}_1,...,\overline{\tau}_M) &=& - \sum_{J=1,...,M} h_J \log(-i\tau_J + i \overline{\tau}_J) \nonumber\\
&+& \sum_{i} \, \frac{\overline{\phi}_i\phi_i}{\displaystyle \prod_{J=1,...,M} (-i \tau_J + i \overline{\tau}_J)^{2k_{i,J}}} 
\,, 
\end{eqnarray}
where all $h_J$ are positive constants. Since each modular symmetry is independent from each other, one modulus field getting a VEV leaves the rest of the K\"ahler potential still satisfying the other modular symmetries. For example, after $\tau_1$ gets a VEV, the K\"ahler potential is left with 
\begin{eqnarray}
- \sum_{J=2,...,M} h_J \log(-i\tau_J + i \overline{\tau}_J) 
+ \sum_{i} \, \frac{1}{\langle -i \tau_J + i \overline{\tau}_J \rangle^{2k_{i,1}}} \frac{\overline{\phi}_i\phi_i}{\displaystyle \prod_{J=2,...,M} (-i \tau_J + i \overline{\tau}_J)^{2k_{i,J}}}  \,. 
\end{eqnarray}
Once all modulus fields get VEVs, the K\"ahler potential gives rise to kinetic terms for the scalar components of the supermultiplets $\phi_i$ and the modulus fields as
\begin{eqnarray}
\sum_{J=1,...,M} \frac{h_J}{\langle -i\tau_J + i \overline{\tau}_J \rangle^2} \partial_\mu \overline{\tau}_J \partial^\mu \tau_J +
\sum_i \frac{\partial_\mu \overline{\phi}_i \partial^\mu \phi_i}{ \displaystyle \prod_{J=1,...,M}\langle -i\tau_J + i \overline{\tau}_J \rangle^{2k_{i,J}} }  \,. 
\end{eqnarray}
In this example, the scalar component of each modulus field performs as a scalar field of vanishing weight in the remaining modular symmetries. 

The superpotential $W(\phi_i;\tau_1,..., \tau_M)$ is in general a function of the modulus fields $\tau_1$ to $\tau_M$ and superfields $\phi_i$. Under the modular transformation, the superpotential should be invariant under the modular transformation \cite{Ferrara:1989bc}. 
Expanding the superpotential $W$ in powers of $\phi_i$, we obtain
\begin{eqnarray}
W(\phi_i;\tau_1,..., \tau_M) = \sum_n  \sum_{\{i_I, \cdots, i_n\}} 
\left(Y_{(I_{Y,1},..., I_{Y,M})} \phi_{i_1} \cdots \phi_{i_n} \right)_{\mathbf{1}} \,, 
\end{eqnarray}
the weights of $Y_{(I_{Y,1}, ..., I_{Y,M})}$ are given by $k_{Y,J} = k_{1,J}+ \cdots k_{n,J}$ for $J=1,...,M$. And
the modular form $Y_{(I_{Y,1}, ..., I_{Y,M})}$ transforms as 
\begin{eqnarray} \label{eq:form_transformation2}
\hspace{-5mm}
&&Y_{(I_{Y,1}, ..., I_{Y,M})}(\tau_1,..., \tau_M) \to Y_{(I_{Y,1},..., I_{Y,M})}(\gamma_1 \tau_1, ..., \gamma_M \tau_M) \nonumber\\
 &&\hspace{2cm}= \prod_{J=1,...,M} (c_J\tau_J + d_J)^{2k_{Y,J}}
 \bigotimes_{J=1,...,M} \rho_{I_{Y,J}}(\gamma_J) Y_{(I_{Y,1},..., I_{Y,M})}(\tau_1,..., \tau_M) \,.
\end{eqnarray} 

\section{Modular $S_4$ symmetry and its remnant symmetries}
\label{S4}
In this section, we temporarily return to the case of a single modular symmetry, focussing on 
the case of a single modular $S_4$ symmetry and its remnant symmetries,
before generalising the results to the case of multiple $S_4$ symmetries in the next section.

\subsection{Modular $S_4$ symmetry}
$S_4$ is a permutation group of four objects. 
In the framework of modular symmetry, the $S_4$ modular group is obtained in the series of $\Gamma_N$ by fixing $N=4$. In other word, its generators satisfy $S_\tau^2  = (S_\tau T_\tau)^3 = T_\tau^4 =I$. 
In previous works, it is common to use three generators $S$, $T$ and $U$, which satisfy $S^2 = T^3 = U^2 = (ST)^3 = (SU)^2 = (TU)^2 =(STU)^4=I$~\cite{King:2013eh}, to generate $S_4$. These traditional generators are related to 
the modular generators $S_\tau$ and $T_\tau$ as
\begin{eqnarray}
S = T_\tau^2 \,,~
T = S_\tau T_\tau \,,~
U = T_\tau S_\tau T_\tau^2 S_\tau \, ,
\end{eqnarray}
which provides a useful dictionary to relate the two types of generators.
In the upper complex plane with the requirement $\tau = \tau +4$, $S$, $T$ and $U$ can be represented by two by two matrices such as
\begin{eqnarray} \label{eq:STU}
S=\begin{pmatrix} 1 & 2 \\ 0 & 1 \end{pmatrix}\,, ~
T=\begin{pmatrix} 0 & 1 \\ -1 & -1 \end{pmatrix} \,, ~
U=\begin{pmatrix} 1 & -1 \\ 2 & -1 \end{pmatrix} \,.
\end{eqnarray}
Due to the identification in Eq.~\eqref{eq:mode_N}, these representation matrices are not unique. 
Using Eq.~\eqref{eq:STU}, we
write out another three elements of $S_4$, namely $TS = S_\tau T_\tau^{-1} $, $ST = T_\tau S_\tau T_\tau^{-1} S_\tau$ and $STS = T_\tau^{-1} S_\tau T_\tau S_\tau$, which are order-three elements of $S_4$ which will appear in our later discussion,
\begin{eqnarray} \label{eq:TS}
TS=\begin{pmatrix} 0 & 1 \\ -1 & 1 \end{pmatrix}\,,~~
ST=\begin{pmatrix} 2 & -1 \\ 3 & -1 \end{pmatrix}\,,~~
STS=\begin{pmatrix} -2 & -1 \\ 3 & 1 \end{pmatrix}\,.
\end{eqnarray}

Modular forms of even weights in a modular $S_4$ symmetry can be explicitly constructed in terms of the Dedekind eta function $\eta(\tau)\equiv q^{1/24} \prod_{n=1}^{\infty} (1- q^n)$, with $q = e^{2\pi i \tau}$ \cite{Penedo:2018nmg}. 
At lowest weight $2k=2$, there are five independent modular forms. 
By defining 
\begin{eqnarray}
Y(a_1, \cdots, a_6 | \tau) &=& \frac{d}{d\tau} \left[ a_1 \log \eta \left( \tau + \frac{1}{2} \right) + a_2 \log \eta \left( 4 \tau \right) + a_3 \log \eta \left( \frac{\tau}{4} \right) \right.\nonumber\\
&&\left.+ a_4 \log \eta \left( \frac{\tau+1}{4} \right) + a_5 \log \eta \left( \frac{\tau+2}{4} \right) + a_6 \log \eta \left( \frac{\tau+3}{4} \right) \right] ,
\end{eqnarray}
with $a_1+\cdots+a_6=0$, these five independent modular forms can be constructed to be
\begin{eqnarray}\label{eq:form}
Y_1(\tau) &=& Y (1, 1, \omega, \omega^2, \omega, \omega^2 | \tau) \,, \nonumber\\
Y_2(\tau) &=& Y (1, 1, \omega^2, \omega, \omega^2, \omega | \tau) \,, \nonumber\\
Y_3(\tau) &=& Y (1, -1, -1, -1, 1, 1 | \tau) \,, \nonumber\\
Y_4(\tau) &=& Y (1, -1, -\omega^2, -\omega, \omega^2, \omega | \tau) \,, \nonumber\\
Y_5(\tau) &=& Y (1, -1, -\omega, -\omega^2, \omega, \omega^2 | \tau) \,, 
\end{eqnarray}
where $\omega= e^{2\pi i/3}$. These five independent modular forms 
at lowest weight $2k=2$ form a doublet $\mathbf{2}$ and a triplet $\mathbf{3}'$ of $S_4$,
\begin{eqnarray} \label{eq:Y2}
Y_{\mathbf{2}}^{(2)} = \begin{pmatrix} Y_1 \\ Y_2 \end{pmatrix}\,, \hspace{1cm}
Y_{\mathbf{3}'}^{(2)} = \begin{pmatrix} Y_3 \\ Y_4 \\ Y_5 \end{pmatrix} \,.
\end{eqnarray}

Modular forms with higher even weights ($2k=4, 6, \cdots$) can be constructed from these five modular forms. In general, the dimension of the linear space formed by the modular forms of weight $2k$ and level 4 is $4k+1$ \cite{Feruglio:2017spp}. Namely, the nine independent modular forms of weight $2k=4$, which form one $\mathbf{1}$, one $\mathbf{2}$, one $\mathbf{3}$ and one $\mathbf{3}'$. Among them, the two triplet modular forms are given by
\begin{eqnarray} \label{eq:Y4}
Y_{\mathbf{3}}^{(4)} = \begin{pmatrix} Y_1 Y_4 - Y_2 Y_5 \\ Y_1 Y_5 - Y_2 Y_3 \\ Y_1 Y_3 - Y_2 Y_4 \end{pmatrix} \,, &\hspace{1cm}&
Y_{\mathbf{3}'}^{(4)} = \begin{pmatrix} Y_1 Y_4 + Y_2 Y_5 \\ Y_1 Y_5 + Y_2 Y_3 \\ Y_1 Y_3 + Y_2 Y_4 \end{pmatrix} \,.
\end{eqnarray}
At weight $2k=6$, there are 13 independent forms. They form one $\mathbf{1}$, one $\mathbf{1}'$, one $\mathbf{2}$, one $\mathbf{3}$ and two $\mathbf{3}'$s of $S_4$. Here we only interested in the two $\mathbf{3}'$s of $S_4$.
They are given by 
\begin{eqnarray} \label{eq:Y6}
Y_{\mathbf{3}}^{(6)} = 
\begin{pmatrix}
 -Y_1^2 Y_5 +Y_2^2 Y_4 \\
 -Y_1^2 Y_3 +Y_2^2 Y_5 \\
 -Y_1^2 Y_4 +Y_2^2 Y_3 
\end{pmatrix} \,, \hspace{1cm}
Y_{\mathbf{3}'_1}^{(6)} = 
\begin{pmatrix}
 Y_1^2 Y_5 +Y_2^2 Y_4 \\
 Y_1^2 Y_3 +Y_2^2 Y_5 \\
 Y_1^2 Y_4 +Y_2^2 Y_3 
\end{pmatrix} \,, \hspace{1cm}
Y_{\mathbf{3}'_2}^{(6)} = 
Y_1 Y_2 \begin{pmatrix}
 Y_3 \\
 Y_4 \\
 Y_5 
\end{pmatrix} \,.
\end{eqnarray} 
These modular forms will be used for our model building in the next section. For modular forms with weights up to 10, a full list can be found in \cite{Novichkov:2018ovf}.

Extension from a single $S_4$ modular symmetry to a series of modular $S_4$ symmetries is straightforwardly achieved by following the procedure in section~\ref{sec:multi_modular} with all levels fixed at $N_J=4$. In each $S_4^J$, we denote their generators $S$, $T$ and $U$ by $S_J$, $T_J$ and $U_J$, where the subscript is only used to distinguish groups. Modular forms with weights $k_{Y,1}, ..., k_{Y,M}$ are multiplets of multiple moduli, namely of of $\tau_1$, ..., $\tau_M$. 

\subsection{Stabilisers and residual symmetries of modular $S_4$\label{sec:residual}}

Although a brief discussion on residual symmetry after modular $S_4$ symmetry breaking has been given in \cite{Novichkov:2018ovf}, we note 
that the essential correlation between the modular field and its residual symmetries has not been discussed. 
In this section, we will give a thorough analysis of this case, uncovering some new results along the way. 

We begin by introducing and reviewing the notion of stabilisers of the symmetry
which will play a crucial role in residual symmetries. 
Given an element $\gamma$ in the modular group $S_4 \simeq \Gamma_4$, 
a stabiliser of $\gamma$ corresponds to a fixed point $\tau_\gamma$ in the upper complex plane which satisfies $\gamma \tau_\gamma = \tau_\gamma$. 
Once the modular field $\tau$ gains a VEV at such a stabiliser, $\langle \tau \rangle = \tau_\gamma$, an Abelian residual modular symmetry generated by $\gamma$ is preserved. 
It is obvious that acting $\gamma$ on a modular form at its stabiliser leaves the modular form invariant, i.e., 
\begin{eqnarray}
\gamma: Y_I(\tau_\gamma) \to Y_I(\gamma \tau_\gamma) = Y_I(\tau_\gamma)\,. 
\end{eqnarray}
Following the standard transformation property in Eq. \eqref{eq:form_transformation}, we obtain 
\begin{eqnarray} \label{eq:yukawa_eigenvector}
\rho_I(\gamma) Y_I(\tau_\gamma) = (c\tau_\gamma + d)^{-2k} Y_I(\tau_\gamma) \,. 
\end{eqnarray} 
This equation lead us to the following important properties for the stabiliser and the modular form:
\begin{itemize}

\item A modular form at a stabiliser $Y_I(\tau_\gamma)$ is an eigenvector of the representation matrix $\rho_I(\gamma)$ with respective eigenvalue $(c\tau_\gamma + d)^{-2k}$.

\item The stabiliser $\tau_\gamma$ satisfies $|c\tau_\gamma + d| = 1$ since $(c\tau_\gamma + d)^{-2k}$ is an eigenvalue of a unitary matrix. 
\end{itemize}
A special case is that when $(c\tau_\gamma + d)^{-2k}=1$ is satisfied, $\rho_I(\gamma) Y_I(\tau_\gamma) = Y_I(\tau_\gamma)$, and we recover the residual flavour symmetry generated by $\gamma$. 
In general, the eigenvalue does not need to be fixed at $1$ in the framework of modular symmetry. 

In the follow-up of this subsection, we will consider the following stabilisers, 
\begin{eqnarray}
&\tau_S= i\infty \,,~
\tau_T= \omega = - \frac{1}{2} + i \frac{\sqrt{3}}{2} \,,~
\tau_U=\frac{1}{2} + \frac{i}{2} \,,\nonumber\\
&\tau_{TS}=-\omega^2 = \frac{1}{2} + i \frac{\sqrt{3}}{2} \,,~
\tau_{ST}=\frac{1}{2}+\frac{i}{2\sqrt{3}}\,,~
\tau_{STS}=-\frac{1}{2}+\frac{i}{2\sqrt{3}} \,.
\end{eqnarray}
Although $\tau_T$ and $\tau_{TS}$ have been discussed in \cite{Novichkov:2018ovf} (identified with $\tau_L$ and $\tau_R$ therein, respectively), $\tau_S$, $\tau_U$ and $\tau_{ST}$ as stabilisers in the $S_4$ modular symmetry are discussed here for the first time. Here we apply this notation to take the advantage of modular residual symmetries generated by $S$, $T$, $U$, $TS$, $ST$ and $STS$, respectively. 
Following Eq.~\eqref{eq:STU}, it is straightforward to check that these stabilisers are invariant under the corresponding modular transformations respectively, i.e., 
\begin{eqnarray}
&&S: \tau_S \to S \tau_S= \tau_S+2 = \tau_S\,, \nonumber\\
&&T: \tau_T \to T \tau_T= \frac{-1}{\tau_T+1} = \tau_T\,, \nonumber\\
&&U: \tau_U \to U \tau_U= \frac{\tau_U-1}{2\tau_U-1} = \tau_U \,, \nonumber\\
&&TS: \tau_{TS} \to TS \, \tau_{TS} = \frac{1}{-\tau_{TS}+1} = \tau_{TS} \,, \nonumber\\
&&ST: \tau_{ST} \to ST \, \tau_{ST} = \frac{2\tau_{ST}-1}{3\tau_{ST}-1} = \tau_{ST} \,, \nonumber\\
&&STS: \tau_{STS} \to STS \, \tau_{STS} = \frac{-2\tau_{STS}-1}{3\tau_{STS}+1} = \tau_{STS} \,. 
\end{eqnarray}
It is worthy noting that these stabilisers are some typical examples but not the full list of stabilisers of $S_4$.

At the stabiliser, the multiplets formed by the modular form may specify interesting directions. 
We will discuss how the triplet modular forms $Y_{\mathbf{3}}^{(2k)}$ or  $Y_{\mathbf{3}'}^{(2k)}$ (for $k =1,2,3$) gain these directions based on the symmetry argument in Eq.~\eqref{eq:yukawa_eigenvector}. 

We begin our discussion from modular forms at the stabiliser $\gamma_S$. 
We know that $Y_{\mathbf{3}^{(\prime)}}^{(2k)}(\tau_S)$ is the eigenvector of $\rho_{\mathbf{3}^{(\prime)}}(S)$ with respective eigenvalue $1^{-2k} \equiv 1$, 
\begin{eqnarray}
\rho_{\mathbf{3}^{(\prime)}}(S) Y^{(2k)}_{\mathbf{3}^{(\prime)}}(\tau_S) = Y^{(2k)}_{\mathbf{3}^{(\prime)}}(\tau_S)
\end{eqnarray}
for any weight $2k$. Given the well-known representation matrix for $S$ in $\mathbf{3}$ or $\mathbf{3}'$, 
\begin{eqnarray}
\rho_{\mathbf{3}^{(\prime)}}(S) = \frac{1}{3} \begin{pmatrix} -1 & 2 & 2 \\ 2 & -1 & 2 \\ 2 & 2 & -1 \end{pmatrix} \,.
\end{eqnarray} 
Three eigenvalues are given by $1$, $-1$ and $-1$. The eigenvector corresponding to the eigenvalue $1$ is always fixed at $(1,1,1)^T$ up to an overall factor. Therefore, we conclude that 
$Y^{(2k)}_{\mathbf{3}^{(\prime)}}(\tau_S)$ always takes the form
\begin{eqnarray} \label{eq:residual_S}
Y^{(2k)}_{\mathbf{3}^{(\prime)}}(\tau_S) = y^{(2k)}_{\mathbf{3}^{(\prime)},S} \begin{pmatrix} 1 \\ 1 \\ 1 \end{pmatrix} \,.
\end{eqnarray}
Here, $y^{(2k)}_{\mathbf{3}^{(\prime)},S}$ is a overall factor determined by the weight and representation. By taking $\tau_S = i \infty$ into the exact modular form $Y_i(\tau)$ in Eq.~\eqref{eq:form}, we obtain $q=0$ and $Y_1(\tau_S) = Y_2(\tau_S) = i 3\pi/8$, $Y_3(\tau_S) = Y_4(\tau_S) = Y_5(\tau_S) = i \pi/4$. For the weight $2k=2$, $y_{\mathbf{3}^{\prime},S}^{(2)}=i \pi/4$ for $\mathbf{3}'$. For $2k=4$, $y_{\mathbf{3},S}^{(4)} = 0$ and $y_{\mathbf{3}^{\prime},S}^{(4)} = -3 \pi^2/16$. 
For $2k=6$, $y_{\mathbf{3},S}^{(6)} = 0$, $y_{\mathbf{3}^{\prime}_1,S}^{(6)} = 2 y_{\mathbf{3}^{\prime}_2,S}^{(6)} = -i 9 \pi^3/128$. 
At the stabiliser $\tau_S$, since the eigenvalue is always fixed at 1 regardless of the weight, the residual $Z_2^S$ modular symmetry is identical to the residual $Z_2^S$ flavour symmetry. 

We perform a similar discussion for modular forms at the stabiliser $\tau_T$. Eq.~\eqref{eq:yukawa_eigenvector} is simplified into 
\begin{eqnarray}
\rho_{\mathbf{3}^{(\prime)}}(T) Y^{(2k)}_{\mathbf{3}^{(\prime)}}(\tau_T) = (-\tau_T-1)^{-2k} Y^{(2k)}_{\mathbf{3}^{(\prime)}}(\tau_T) = \omega^{2k} Y^{(2k)}_{\mathbf{3}^{(\prime)}}(\tau_T) \,.
\end{eqnarray}
Thus, the selected eigenvector corresponds to the eigenvalue $\omega^{2k}$, which is weight-dependent. In the $T$-diagonal basis we use in the paper, representation matrix for $T$ is given by 
\begin{eqnarray}
\rho_{\mathbf{3}^{(\prime)}}(T)=\begin{pmatrix} 1 & 0 & 0 \\ 0 & \omega^2 & 0 \\ 0 & 0 & \omega \end{pmatrix} \,.
\end{eqnarray} 
The triplet form, as an eigenvalue of $T$, takes a very simple form
\begin{eqnarray} \label{eq:residual_T}
&& Y^{(2)}_{\mathbf{3}'}(\tau_T) = y_{\mathbf{3}',T}^{(2)} \begin{pmatrix} 0 \\ 1 \\ 0 \end{pmatrix}\,,~
Y^{(4)}_{\mathbf{3}^{(\prime)}}(\tau_T) = y_{\mathbf{3}^{(\prime)},T}^{(4)} \begin{pmatrix} 0 \\ 0 \\ 1 \end{pmatrix}\,,~
Y^{(6)}_{\mathbf{3}^{(\prime)}}(\tau_T) = y_{\mathbf{3}^{(\prime)},T}^{(6)} \begin{pmatrix} 1 \\ 0 \\ 0 \end{pmatrix}\,, 
\end{eqnarray}
where the overall factors are also determined by the weight and representation.
These results can be checked numerically by taking $\tau_T$ into exact formulas of modular forms. It is straightforward to obtain $Y_1(\tau_T) = Y_3(\tau_T) = Y_5(\tau_T) = 0$, and we are left with only two non-zero modular forms, $Y_2(\tau_T) = 2.11219 i$ and $Y_4(\tau_T) = -2.43895 i$. Taking them to Eqs.~\eqref{eq:Y2}, \eqref{eq:Y4} and \eqref{eq:Y6}, we arrive at the same above result with $y_{\mathbf{3}',T}^{(2)} = -2.43895 i$, $y_{\mathbf{3},T}^{(4)} = - y_{\mathbf{3}',T}^{(4)} = -5.15151$, and $y_{\mathbf{3},T}^{(6)} = y_{\mathbf{3}'_1,T}^{(6)} = 10.881 i$,  and $y^{(6)}_{\mathbf{3}'_2,T} = 0$. In this typical example, only the third direction, i.e., $(1,0,0)^T$, corresponding to modular forms with weights $2k = 0~ ({\rm mod} ~3)$, preserves the residual flavour symmetry generated by $T$. The other two vectors do not satisfy the residual flavour symmetry, but only the residual modular symmetry. 

In the framework of flavour symmetry, the residual symmetry generated by $U$ is usually called $\mu$-$\tau$ symmetry. We discuss the modular form at the stabiliser of $U$. 
$Y_{\mathbf{3}^{(\prime)}}^{(2k)}(\tau_U)$ is the eigenvalue of $\rho_{\mathbf{3}^{(\prime)}}(U)$ with respective eigenvalue $(2\tau_U-1)^{-2k} = (-1)^{k}$. Representation matrices for $U$ are different in $\mathbf{3}$ and $\mathbf{3}'$, 
\begin{eqnarray}
\rho_{\mathbf{3}}(U) = \begin{pmatrix} 1 & 0 & 0 \\ 0 & 0 & 1 \\ 0 & 1 & 0 \end{pmatrix} \,,~
\rho_{\mathbf{3}'}(U) = - \begin{pmatrix} 1 & 0 & 0 \\ 0 & 0 & 1 \\ 0 & 1 & 0 \end{pmatrix} \,.
\end{eqnarray} 
$\rho_{\mathbf{3}}(U)$ has one eigenvalue $-1$ and the other two degenerate eigenvalues $+1$. The eigenvector with respective eigenvalue $-1$ is fixed at $(0,1,-1)^T$ without considering an overall factor. The eigenvector with respective eigenvalue $+1$ is in principle a linear combination of two independent vectors $(2,-1,-1)^T$ and $(1,1,1)^T$. For odd and even $k$, we can express $Y_{\mathbf{3}}^{(2k)}(\tau_U)$ as 
\begin{eqnarray} \label{eq:residual_U}
&&Y^{(2k)}_{\mathbf{3}}(\tau_U) = y^{(2k)}_{\mathbf{3},U} \begin{pmatrix} 0 \\ 1 \\ -1  \end{pmatrix}~ \hspace{23mm} \text{for an odd } k\,, \nonumber\\
&&Y^{(2k)}_{\mathbf{3}}(\tau_U) = y^{(2k)}_{\mathbf{3},U} \begin{pmatrix} 2 \\ -1 \\ -1  \end{pmatrix}
+ y^{(2k)\prime}_{\mathbf{3},U} \begin{pmatrix} 1 \\ 1 \\ 1 \end{pmatrix}~ \text{for an even } k\,.
\end{eqnarray}
The coefficients are determined by the weight. 
Numerically, $Y_1(\tau_U)=-Y_2(\tau_U)=2.84287 i$, $Y_3(\tau_U)=-(2\sqrt{2}+i)a$, $Y_4(\tau_U)=Y_5(\tau_U)=(\sqrt{2}-i)a$ with $a=1.09422$. We obtain $y^{(4)}_{\mathbf{3},U}=\sqrt{2}Y_1(\tau_U) a$, $y^{(4)\prime}_{\mathbf{3},U} = - i 2 Y_1(\tau_U) a$ for $2k=4$, and $y^{(6)}_{\mathbf{3},U}=3\sqrt{2}Y_1^2(\tau_U) a$ for $2k=6$.  In $\mathbf{3}'$ representations, $\rho_{\mathbf{3}'}(U)$ has one eigenvalue $+1$ and the other two degenerate eigenvalues $-1$. For an even $k$ the direction of $Y^{(2k)}_{\mathbf{3}'}(\tau_U)$ is fixed along $(0, 1, -1)^T$, while for an odd $k$ $Y^{(2k)}_{\mathbf{3}'}(\tau_U)$ is a linear combination of $(2,-1,-1)^T$ and $(1,1,1)^T$, 
\begin{eqnarray} \label{eq:residual_U_prime}
&&Y^{(2k)}_{\mathbf{3}'}(\tau_U) = y^{(2k)}_{\mathbf{3}',U} \begin{pmatrix} 2 \\ -1 \\ -1  \end{pmatrix}
+ y^{(2k)\prime}_{\mathbf{3}',U} \begin{pmatrix} 1 \\ 1 \\ 1 \end{pmatrix}~ \text{for an odd } k\,, \nonumber\\
&&Y^{(2k)}_{\mathbf{3}'}(\tau_U) = y^{(2k)}_{\mathbf{3}',U} \begin{pmatrix} 0 \\ 1 \\ -1  \end{pmatrix}~ \hspace{23mm} \text{for an even } k\,.
\end{eqnarray}
Specifically, for $2k=2$, we have 
$y_{\mathbf{3}',U}^{(2)} = - \sqrt{2} a$, 
$y_{\mathbf{3}',U}^{(2)\prime} = - i a$.
For $2k=6$, we have
$y^{(6)}_{\mathbf{3}_1,U}=\sqrt{2}Y_1^2(\tau_U) a$, 
$y^{(6)\prime}_{\mathbf{3}_1,U} = - i 2 Y_1^2(\tau_U) a$; 
and 
$y_{\mathbf{3}',U}^{(6)} = \sqrt{2} Y_1^2(\tau) a$, 
$y_{\mathbf{3}',U}^{(6)\prime} = i Y_1^2(\tau) a$, 
respectively. 
And for $2k=4$, keeping the $(0,1,-1)^T$, we have $y_{\mathbf{3}',U}^{(4)} = 3\sqrt{2} Y_1(\tau_U) a$.
We would like to mention that although the direction $(0,1,-1)^T$ is realised in both $\mathbf{3}$ and $\mathbf{3}'$ representations, $(0,1,-1)^T$ in $\mathbf{3}'$ preserves a $\mu$-$\tau$ flavour symmetry, but that in $\mathbf{3}$ preserves not a $\mu$-$\tau$ flavour symmetry, but a $\mu$-$\tau$ modular symmetry.

In addition, we would like to consider stabilisers for the elements $TS$, $ST$ and $STS$. These elements are order-3 elements and stabiliser for each element preserves a $Z_3$ symmetry. The representation matrices of $TS$, $ST$ and $STS$ take the forms
\begin{eqnarray}
\rho_{\mathbf{3}^{(\prime)}}(TS) = \frac{1}{3} \begin{pmatrix} -1 & 2 & 2 \\ 2\omega^2 & -\omega^2 & 2\omega^2 \\ 2\omega & 2\omega & -\omega \end{pmatrix} \,,\nonumber\\
\rho_{\mathbf{3}^{(\prime)}}(ST) = \frac{1}{3} \begin{pmatrix} -1 & 2\omega^2 & 2\omega \\ 2 & -\omega^2 & 2\omega \\ 2 & 2\omega^2 & -\omega \end{pmatrix} \,,\nonumber\\
\rho_{\mathbf{3}^{(\prime)}}(STS) = \frac{1}{3} \begin{pmatrix} -1 & 2\omega & 2\omega^2 \\ 2\omega & -\omega^2 & 2 \\ 2\omega^2 & 2 & -\omega \end{pmatrix} \,.
\end{eqnarray}
They all have three eigenvalues given by $1$, $\omega$ and $\omega^2$. The corresponding eigenvectors for $TS$ are $(-1, 2\omega, 2\omega^2)^T$, $(2\omega, 2\omega^2, -1)^T$ and $(2\omega^2,-1,2\omega)^T$, respectively; the corresponding eigenvectors for $ST$ are $(-1, 2\omega^2, 2\omega)^T$, $(2\omega^2, 2\omega, -1)^T$ and $(2\omega,-1,2\omega^2)^T$, respectively; and the corresponding eigenvectors for $STS$ are $(-1, 2, 2)^T$, $(2, 2, -1)^T$ and $(2,-1,2)^T$, respectively. 
$Y_{\mathbf{3}^{(\prime)}}^{(2k)}(\tau_{TS})$ corresponds to the eigenvalue $(1-\tau_{TS})^{-2k} = \omega^{k}$. Thus, we directly arrive at
\begin{eqnarray} \label{eq:residual_STT}
Y^{(2)}_{\mathbf{3}'}(\tau_{TS}) = y^{(2)}_{\mathbf{3}',TS} \begin{pmatrix} 2\omega \\ 2\omega^2 \\ -1 \end{pmatrix},\,
Y^{(4)}_{\mathbf{3}^{(\prime)}}(\tau_{TS}) = y^{(4)}_{\mathbf{3}^{(\prime)},TS} \begin{pmatrix} 2\omega^2 \\ -1 \\ 2\omega \end{pmatrix},\,
Y^{(6)}_{\mathbf{3}}(\tau_{TS}) = y^{(6)}_{\mathbf{3}^{(\prime)},TS} \begin{pmatrix} -1 \\ 2\omega \\ 2\omega^2 \end{pmatrix}.
\end{eqnarray}
Taking the explicit formulas of modular forms into account, we obtain the overall factors to be $y^{(2)}_{\mathbf{3}',TS} = - Y_5(\tau_{TS}) = 0.81298i$, $y^{(4)}_{\mathbf{3}, TS} = y^{(4)}_{\mathbf{3}', TS} = -Y_1(\tau_{TS}) Y_5(\tau_{TS}) = - 1.71717$, $y^{(6)}_{\mathbf{3},TS} = - y^{(6)}_{\mathbf{3}'_1,TS} = -Y_1^2(\tau_{TS}) Y_5(\tau_{TS}) = - 3.62699 i$, and $y^{(6)}_{\mathbf{3}'_2,TS} = 0$. We turn to the modular forms at stabilisers $\tau_{ST}$. $Y_{\mathbf{3}^{(\prime)}}^{(2k)}(\tau_{ST})$ are obtained by exchanging the second and the third entries of the above expressions but with care due to different weights
\begin{eqnarray} \label{eq:residual_STT}
Y^{(2)}_{\mathbf{3}'}(\tau_{ST}) = y^{(2)}_{\mathbf{3}',ST} \begin{pmatrix} 2\omega^2 \\ -1 \\ 2\omega \end{pmatrix},\,
Y^{(4)}_{\mathbf{3}^{(\prime)}}(\tau_{ST}) = y^{(4)}_{\mathbf{3}^{(\prime)},ST} \begin{pmatrix} 2\omega \\ 2\omega^2 \\ -1 \end{pmatrix},\,
Y^{(6)}_{\mathbf{3}}(\tau_{ST}) = y^{(6)}_{\mathbf{3}^{(\prime)},ST} \begin{pmatrix} -1 \\ 2\omega \\ 2\omega^2 \end{pmatrix},
\end{eqnarray}
where $y^{(2)}_{\mathbf{3}',ST} = 2.43895 i$, $y^{(4)}_{\mathbf{3}, ST} = -y^{(4)}_{\mathbf{3}', ST}  = -15.4545$,  $y^{(6)}_{\mathbf{3}, ST} = y^{(6)}_{\mathbf{3}'_1, ST} = -97.9287 i$ and $y^{(6)}_{\mathbf{3}'_2, ST} = 0$. 
They correspond to eigenvectors of $\rho_{\mathbf{3}^{(\prime)}}(ST) $ with respective eigenvalues $(3\tau_{ST}-1)^{-2k} = \omega^{2k}$. 
Finally, we list modular forms at stabilisers $\tau_{STS}$. $Y_{\mathbf{3}^{(\prime)}}^{(2k)}(\tau_{STS})$ are given by 
\begin{eqnarray} \label{eq:residual_STT}
Y^{(2)}_{\mathbf{3}'}(\tau_{STS}) \!=\! y^{(2)}_{\mathbf{3}', ST} \begin{pmatrix} 2 \\ 2 \\ -1 \end{pmatrix},\,
Y^{(4)}_{\mathbf{3}^{(\prime)}}(\tau_{STS}) \!=\! y^{(4)}_{\mathbf{3}^{(\prime)}, STS} \begin{pmatrix} 2 \\ -1 \\ 2 \end{pmatrix},\,
Y^{(6)}_{\mathbf{3}}(\tau_{STS}) \!=\! y^{(6)}_{\mathbf{3}^{(\prime)}, STS} \begin{pmatrix} -1 \\ 2 \\ 2 \end{pmatrix},
\end{eqnarray}
where $y^{(2)}_{\mathbf{3}', STS} = - 2.43895 i$, $y^{(4)}_{\mathbf{3}, STS} = y^{(4)}_{\mathbf{3}', STS} = -15.4545$, $y^{(6)}_{\mathbf{3},STS} = - y^{(6)}_{\mathbf{3}'_1,STS} = -97.9287 i$, and $y^{(6)}_{\mathbf{3}'_2,STS} = 0$. 
They correspond to eigenvectors of $\rho_{\mathbf{3}^{(\prime)}}(ST) $ with respective eigenvalues $(3\tau_{STS}+1)^{-2k} = \omega^{k}$. 

We summarise directions of triplet ($\mathbf{3}$ and $\mathbf{3}'$) modular forms for lower weights ($2k=2,4,6$) at stabilisers ($\tau = \tau_S, \tau_U, \tau_T, \tau_{TS}, \tau_{ST}, \tau_{STS})$ in Table~\ref{tab:stabilisers}. 

All the above discussion in this subsection is based on a single modular $S_4$ with a single modulus field. Extending to the case of multiple modular symmetries may allow the theory to have several different residual modular symmetries. Namely, the different moduli fields may take different values at different stabilisers. In the next section, we will apply this property to model building.

%%%%%%%%%%%%%%%%%%%%%%%%%%%%%%%%%%%%%%%%%%%%%%%%
\begin{table}[h]
\begin{center}
\begin{tabular}{| c | c | c | c | c | c | c | c |}
\hline \hline
\multirow{2}{*}{$\tau$} 
& weight 2 & \multicolumn{2}{c|}{weight 4} & \multicolumn{3}{c|}{weight 6} \\\cline{2-7} 
& $\mathbf{3}'$ & $\mathbf{3}$ & $\mathbf{3}'$ & $\mathbf{3}$ & $\mathbf{3}'_1$ & $\mathbf{3}'_2$ \\
\hline \hline

$\tau_S$ & $\begin{pmatrix} 1 \\ 1 \\ 1 \end{pmatrix}$ & 
$\begin{pmatrix} 1 \\ 1 \\ 1 \end{pmatrix}$ & 
$\begin{pmatrix} 1 \\ 1 \\ 1 \end{pmatrix}$ &
$\mathbf{0}$ 
& $\begin{pmatrix} 1 \\ 1 \\ 1 \end{pmatrix}$ &
$\begin{pmatrix} 1 \\ 1 \\ 1 \end{pmatrix}$ \\\hline

$\tau_U$ & $\begin{pmatrix} 0 \\ 1 \\ -1 \end{pmatrix}$ & 
$ \begin{pmatrix} 2-i\sqrt{2} \\ -1-i\sqrt{2} \\ -1-i\sqrt{2}  \end{pmatrix}$ & 
$\begin{pmatrix} 0 \\ 1 \\ -1 \end{pmatrix}$ &
$\begin{pmatrix} 0 \\ 1 \\ -1 \end{pmatrix}$ &
$ \begin{pmatrix} 2\sqrt{2} + i \\ -\sqrt{2} + i \\ -\sqrt{2} + i  \end{pmatrix}$ & 
$ \begin{pmatrix} 2-i\sqrt{2} \\ -1-i\sqrt{2} \\ -1-i\sqrt{2}  \end{pmatrix}$ 
 \\\hline

$\tau_T$ & $\begin{pmatrix} 0 \\ 1 \\ 0 \end{pmatrix}$ &
$\begin{pmatrix} 0 \\ 0 \\ 1 \end{pmatrix}$ &
$\begin{pmatrix} 0 \\ 0 \\ 1 \end{pmatrix}$ &
$\begin{pmatrix} 1 \\ 0 \\ 0 \end{pmatrix}$ &
$\begin{pmatrix} 1 \\ 0 \\ 0 \end{pmatrix}$ & $\mathbf{0}$ \\\hline

$\tau_{TS}$ & 
$\begin{pmatrix} 2\omega \\ 2\omega^2 \\ -1 \end{pmatrix}$ &
$\begin{pmatrix} 2\omega^2 \\ -1 \\ 2\omega \end{pmatrix}$ &
$\begin{pmatrix} 2\omega^2 \\ -1 \\ 2\omega \end{pmatrix}$ &
$\begin{pmatrix} -1 \\ 2\omega \\ 2\omega^2 \end{pmatrix}$ &
$\begin{pmatrix} -1 \\ 2\omega \\ 2\omega^2 \end{pmatrix}$ & $\mathbf{0}$ \\\hline

$\tau_{ST}$ & 
$\begin{pmatrix} 2\omega \\ -1 \\ 2\omega^2 \end{pmatrix}$ &
$\begin{pmatrix} 2\omega^2 \\ 2\omega \\ -1 \end{pmatrix}$ &
$\begin{pmatrix} 2\omega^2 \\ 2\omega \\ -1 \end{pmatrix}$ &
$\begin{pmatrix} -1 \\ 2\omega^2 \\ 2\omega \end{pmatrix}$ &
$\begin{pmatrix} -1 \\ 2\omega^2 \\ 2\omega \end{pmatrix}$ & $\mathbf{0}$ \\\hline
$\tau_{STS}$ & 
$\begin{pmatrix} 2 \\ 2 \\ -1 \end{pmatrix}$ &
$\begin{pmatrix} 2 \\ -1 \\ 2 \end{pmatrix}$ &
$\begin{pmatrix} 2 \\ -1 \\ 2 \end{pmatrix}$ &
$\begin{pmatrix} -1 \\ 2 \\ 2 \end{pmatrix}$ &
$\begin{pmatrix} -1 \\ 2 \\ 2 \end{pmatrix}$ & $\mathbf{0}$ \\
\hline \hline
\end{tabular}
\end{center}
\caption{Triplet ($\mathbf{3}$ and $\mathbf{3}'$) representations of $S_4$ modular forms of low weights ($2k= 2, 4, 6$) at typical stabilisers $\tau_S= i\infty$, $\tau_U=\frac{1}{2} + \frac{i}{2}$, $\tau_T= -\frac{1}{2} + i\frac{\sqrt{3}}{2}$, $\tau_{TS}=\frac{1}{2} + i\frac{\sqrt{3}}{2}$, $\tau_{ST}=\frac{1}{2} + \frac{i}{2\sqrt{3}}$ and $\tau_{STS}=-\frac{1}{2} + \frac{i}{2\sqrt{3}}$. Here, we have ignored the overall factor if it is non-zero. $\mathbf{0}$ represents a vanishing modular form, namely, the one which has a zero overall factor. }
 \label{tab:stabilisers}
\end{table}
%%%%%%%%%%%%%%%%%%%%%%%%%%%%%%%%%%%%%%%%%%%%%%%%

\section{A model with three modular $S_4$ symmetries}
\label{model}

\begin{table}[h] \begin{tabular}{| l | c c c c c c|}
\hline \hline
Field & $S_4^A$ & $S_4^B$ & $S_4^C$ & \!$2k_A$\! & \!$2k_B$\! & \!$2k_C$\!\\ 
\hline \hline
$L$ & $\mathbf{1}$ & $\mathbf{1}$ & $\mathbf{3}$ & 0 & 0 & 0\\
$e^c$ & $\mathbf{1}$ & $\mathbf{1}$ & $\mathbf{1}$ & 0 & 0 & \!$-6$\! \\
$\mu^c$ & $\mathbf{1}$ & $\mathbf{1}$ & $\mathbf{1}$ & 0 & 0 & \!$-4$\! \\
$\tau^c$ & $\mathbf{1}$ & $\mathbf{1}$ & $\mathbf{1}$ & 0 & 0 & \!$-2$\! \\
$N_A^c$ & $\mathbf{1}$ & $\mathbf{1}$ & $\mathbf{1}$ & \!$-6$\! & 0 & 0 \\
$N_B^c$ & $\mathbf{1}$ & $\mathbf{1}$ & $\mathbf{1}$ & 0 & \!$-4$\! & 0 \\
\hline 
$\Phi_{AC}$ & $\mathbf{3}$ & $\mathbf{1}$ & $\mathbf{3}$ & 0 & 0 & 0 \\
$\Phi_{BC}$ & $\mathbf{1}$ & $\mathbf{3}$ & $\mathbf{3}$ & 0 & 0 & 0 \\
\hline \hline
%\multicolumn{7}{c}{}\\
\end{tabular}
\begin{tabular}{| l | c c c c c c|}
\hline \hline
Yuk/Mass &$S_4^A$ & $S_4^B$ & $S_4^C$ & \!$2k_A$\! & \!$2k_B$\! & \!$2k_C$\!\\
%/ masses &&&&&& \\ 
\hline \hline
$Y_e(\tau_C)$ & $\mathbf{1}$ & $\mathbf{1}$ & $\mathbf{3}$ & 0 & 0 & $6$ \\
$Y_\mu(\tau_C)$ & $\mathbf{1}$ & $\mathbf{1}$ & $\mathbf{3}$ & 0 & 0 & $4$ \\
$Y_\tau(\tau_C)$ & $\mathbf{1}$ & $\mathbf{1}$ & $\mathbf{3}$ & 0 & 0 & $2$ \\
$Y_A(\tau_A)$ & $\mathbf{3}$ & $\mathbf{1}$ & $\mathbf{1}$ & $6$ & 0 & 0 \\
$Y_B(\tau_B)$ & $\mathbf{1}$ & $\mathbf{3}$ & $\mathbf{1}$ & 0 & $4$ & 0 \\\hline
$M_A(\tau_A)$ & $\mathbf{1}$ & $\mathbf{1}$ & $\mathbf{1}$ & $12$ & 0 & 0 \\
$M_B(\tau_B)$ & $\mathbf{1}$ & $\mathbf{1}$ & $\mathbf{1}$ & 0 & $8$ & 0 
\\
$M_{AB}(\tau_A,\tau_B)$ & $\mathbf{1}$ & $\mathbf{1}$ & $\mathbf{1}$ & $6$ & $4$ & 0 
\\
\hline \hline
\end{tabular}
\caption{Transformation properties of leptons, Yukawa couplings $Y$ and right-handed neutrino masses 
$M$ in $S_4^A \times S_4^B \times S_4^C$. }
\label{tab:particle_contents}
\end{table}

Combining the results of the previous two sections, we see that 
the extension from one single modular field to multiple moduli fields, as discussed in section~\ref{multiple},
opens a window into a new type of modular model building, in which several moduli fields can appear, 
with each one having a different modular form with a different residual symmetry, of the kind discussed in
section~\ref{S4}.

\subsection{A modular $S_4^3$ model}

As a concrete example, we will show how the results of the previous sections can lead to a consistent model of 
trimaximal TM$_1$ mixing, analogous to the traditional approach~\cite{Varzielas:2012pa, Luhn:2013vna}.
At high energies, the model in Table~\ref{tab:particle_contents} is based on three modular symmetries, $S_4^A$, $S_4^B$ and $S_4^C$, with moduli fields labelled by 
$\tau_A$, $\tau_B$ and $\tau_C$, respectively. After the moduli fields gain different VEVs, different textures of mass matrices are realised in charged lepton and neutrino sectors. 

The transformation properties of the leptons are given in Table~\ref{tab:particle_contents}. We arrange that each lepton has no more than one non-vanishing modular weight in either $S_4^A$, $S_4^B$ or $S_4^C$.
We note that: 1) The lepton doublets $L$ form a triplet of $S_4^C$ with zero weight; 2) the right-handed leptons $e^c$, $\mu^c$ and $\tau^c$ are singlets of $S_4^C$ but have different weights $2k_C=-6,-4,-2$, respectively; 3) We introduce only two right-handed neutrinos $N_A^c$ and $N_B^c$, which are all singlets but have weights $2k_A=-6$ and $2k_B=-4$ in $S_4^A$ and $S_4^B$, respectively. It is in principle possible to arrange one field with non-vanishing weights in more than one modular symmetry, so our choice here is just for simplicity. 

In addition, we introduce two scalars $\Phi_{AC}$ and $\Phi_{BC}$. These scalars are assumed to be bi-triplets in the flavour space, arranged in $S_4^A \times S_4^B \times S_4^C$ as $\Phi_{AC}\sim (\mathbf{3}, \mathbf{1}, \mathbf{3})$ and $\Phi_{BC} \sim (\mathbf{1}, \mathbf{3}, \mathbf{3})$ with zero weights. As bi-triplets, they transform as
\begin{eqnarray}
\Phi_{AC} &\to& \rho_{\mathbf{3}}(\gamma_A) \otimes \rho_{\mathbf{3}}(\gamma_C) \Phi_{AC} \,,\nonumber\\
\Phi_{BC} &\to& \rho_{\mathbf{3}}(\gamma_B) \otimes \rho_{\mathbf{3}}(\gamma_C) \Phi_{BC} \,.
\end{eqnarray}
for any elements $\gamma_A$, $\gamma_B$ and $\gamma_C$ of $S_4^A$, $S_4^B$ and $S_4^C$, respectively. 
These scalars are introduced to connect three $S_4$'s together as shown in the superpotential below,
\begin{eqnarray}
w_\ell &=&
\frac{1}{\Lambda}\left[L \Phi_{AC} Y_A(\tau_A) N_A^c + L \Phi_{BC} Y_B(\tau_B) N_B^c \right] H_u \nonumber\\
&&+ \left[ L Y_e(\tau_C) e^c + L Y_\mu(\tau_C) \mu^c + L Y_\tau(\tau_C) \tau^c \right] H_d \nonumber\\
&&+ \frac{1}{2} M_A(\tau_A) N_A^c N_A^c + \frac{1}{2} M_B(\tau_B) N_B^c N_B^c + M_{AB}(\tau_A,\tau_B) N_A^c N_B^c\,, 
\end{eqnarray}
where the leptonic superpotential includes the terms responsible for generating lepton masses.

To be invariant under the modular transformation, 
$Y_{e,\mu,\tau}$ are $\mathbf{3}$-plet modular forms in the modular space $S_4^C$ with weight $2k_C=2,4,6$, respectively, $Y_A$ and $Y_B$ are $\mathbf{3}$-plet modular forms in the modular space $S_4^A$, $S_4^B$ with weights $2k_A = 6$ and $2k_B = 4$, respectively. A term, e.g., $L \Phi_{AC} Y_A(\tau_A) N_A^c$ is explicitly written as 
\begin{eqnarray} \label{eq:example}
L \Phi_{AC} Y_A(\tau_A) N_A^c 
&=& L_1\left[ (\Phi_{AC})_{11} (Y_A)_1 + (\Phi_{AC})_{21} (Y_A)_3 + (\Phi_{AC})_{31} (Y_A)_2 \right] N_A^c \nonumber\\
&+& L_2 \left[ (\Phi_{AC})_{13} (Y_A)_1 + (\Phi_{AC})_{23} (Y_A)_3 + (\Phi_{AC})_{33} (Y_A)_2 \right] N_A^c  \nonumber\\
&+& L_3 \left[ (\Phi_{AC})_{12} (Y_A)_1 + (\Phi_{AC})_{22} (Y_A)_3 + (\Phi_{AC})_{32} (Y_A)_2 \right] N_A^c \,,\nonumber\\
&=& (L_1, L_2, L_3) P_{23} \begin{pmatrix} 
(\Phi_{AC})_{11} & (\Phi_{AC})_{12} & (\Phi_{AC})_{13} \\
(\Phi_{AC})_{21} & (\Phi_{AC})_{22} & (\Phi_{AC})_{23} \\ 
(\Phi_{AC})_{31} & (\Phi_{AC})_{32} & (\Phi_{AC})_{33} \end{pmatrix}^T P_{23}
\begin{pmatrix} (Y_A)_1 \\ (Y_A)_2 \\ (Y_A)_3 \end{pmatrix} 
N_A^c\,, \nonumber\\
\end{eqnarray}
where $L_\alpha$, $(\Phi_{AC})_{i\alpha}$ and $(Y_A)_i$ are entries of $L$, $\Phi_{AC}$ and $Y_A$, respectively, for $i, \alpha=1,2,3$, and $P_{23}$ is the (2,3) row/column-switching transformation matrix. 
$M_A$ and $M_B$ are singlet modular forms in the modular space $S_4^A$, $S_4^B$ with weights $2k_A = 12$ and $2k_B = 8$, respectively. The cross mass term between $N_A$ and $N_B$, $M_{AB}$, is not forbidden. It takes both non-trivial weights in $S_4^A$ and $S_4^B$, $2k_A = 6$ and $2k_B = 4$. The general formulae for $M_A$, $M_B$ and $M_{AB}$ are given by
\begin{eqnarray}
M_A(\tau_A) &=& m_A Y_1^2(\tau_A) Y_2^2(\tau_A) \,, \nonumber\\
M_B(\tau_B) &=& m_{B,1} [Y_1^6(\tau_B) + Y_2^6(\tau_B)] + m_{B,2} Y_1^3(\tau_B) Y_2^3(\tau_B) \,,\nonumber\\
M_{AB}(\tau_A,\tau_B) &=& m_{AB}  [Y_1^3(\tau_A) + Y_2^3(\tau_A)] Y_1(\tau_B) Y_2(\tau_B) \,,
\end{eqnarray}
where $m_A$, $m_{B,1}$, $m_{B,2}$ and $m_{AB}$ are complex free parameters with a mass dimension. 

\subsection{Symmetry breaking of $S_4^3$ to the diagonal $S_4$ subgroup}

\begin{figure}[ht]
\centering

\hspace*{1ex}
	\includegraphics[width=0.5\textwidth]{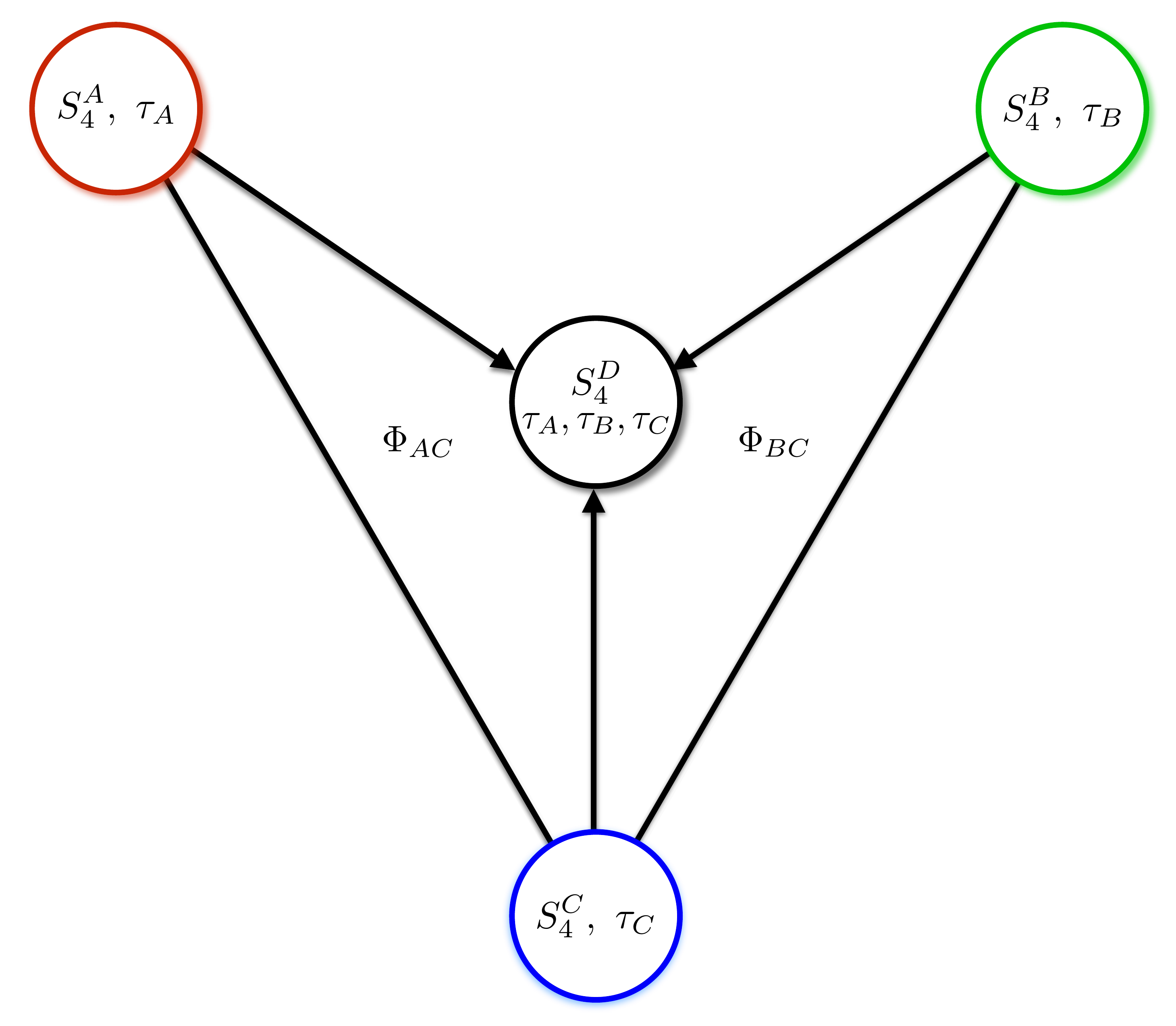}
\caption{Illustration of the breaking of $S_4^A \times S_4^B \times S_4^C \to S_4^D$, identified as the diagonal subgroup, via the VEVs of $\Phi_{AC}$ and $\Phi_{BC}$.}
\label{fig:S4s}
\end{figure}

The modular symmetries are broken after the bi-triplet scalars $\tau_A$, $\tau_B$ and $\tau_C$ gain VEVs. 
Unlike the flavons introduced in most flavour models in the literature, the VEVs of these scalars are not
responsible for special Yukawa textures for leptons, but rather their purpose is to break three modular $S_4$'s to a single modular $S_4$ symmetry, identified as the diagonal subgroup and denoted as $S_4^D$,
\begin{eqnarray}
S_4^A \times S_4^B \times S_4^C \to S_4^D\,,
\end{eqnarray}
as depicted in Fig.~\ref{fig:S4s}.

The VEVs of $\Phi_{AC}$ and $\Phi_{BC}$ take the following forms
\begin{eqnarray} \label{eq:vev}
\langle \Phi_{AC} \rangle_{i \alpha} = v_{AC} (P_{23})_{i \alpha}\,,~~
\langle \Phi_{BC} \rangle_{m \alpha} = v_{BC} (P_{23})_{m \alpha}\,.
\end{eqnarray}
Here again, $P_{23}$ represents the (2,3) row/column-switching transformation matrix, and $\alpha=1,2,3$ corresponds the entries of the triplet of $S_4^C$, while $i=1,2,3$ ($m=1,2,3$) corresponds to those of $S_4^A$ ($S_4^B$). 
These VEV structures are not arbitrarily assumed, but can be simply achieved following the standard driving field method. They are essentially related to the group structure of $S_4$ and its explicit form is basis-dependent \footnote{Explicit forms of scalar VEVs are dependent upon the basis of $S_4$ we use. As shown in Appendix~\ref{app:S4}, we work in the $T$-diagonal basis in Table~\ref{tab:rep_matrix_main}, where the trivial singlet contraction for two triplets is $(ab)_{\mathbf{1}} = a_1 b_1 + a_2 b_3 + a_3 b_2$. If we had worked in the real basis in Table~\ref{tab:rep_matrix_vacuum}, where the singlet contraction can be simply given by $(\tilde{a}\tilde{b})_{\mathbf{1}} = \tilde{a}_1 \tilde{b}_1 + \tilde{a}_2 \tilde{b}_2 + \tilde{a}_3 \tilde{b}_3$, the VEVs of $\Phi_{AC}$ and $\Phi_{BC}$ would have been proportional to the identity matrix, $\langle \tilde{\Phi}_{AC} \rangle_{i \alpha} = v_{AC} \delta_{i\alpha}$, $\langle \tilde{\Phi}_{BC} \rangle_{m \alpha} = v_{BC} \delta_{m \alpha}$, following the discussion in Appendix~\ref{eq:vacuum}. }. For details of how to derive them without loss of generality, we refer the reader to Appendix~\ref{eq:vacuum}. 

Although $S_4^A$, $S_4^B$ and $S_4^C$ are broken by these VEVs,
the diagonal subgroup $S_4^D$ survives below the symmetry breaking scale, 
corresponding to the associated transformation 
$\gamma_A=\gamma_B=\gamma_C$. 
In more detail, the $S_4^D$ survives since, given any $\gamma_A$ of $S_4^A$, there always exists an element $\gamma_C$ of $S_4^C$ which is identical to $\gamma_A$, and the VEV of $\Phi_{AC}$ is invariant under this ``contravariant'' transformation. Furthermore, there also exists an element $\gamma_B$ of $S_4^B$ which is identical to $\gamma_C$, and the VEV of $\Phi_{BC}$ is also invariant 
under the transformation. Thus, the modular $S_4^D$ symmetry corresponds to a universal transformation. 

\subsection{The effective low energy theory with modular $S_4$ symmetry}

The effective low energy superpotential, below the $S_4^3$ breaking scale, involves only a single 
surviving modular $S_4$ symmetry,  and may be written as, 
\begin{eqnarray} 
w^{\rm eff}_\ell \!&=&\! 
\left[ \frac{v_{AC}}{\Lambda} L Y_A(\tau_A) N_A^c + \frac{v_{BC}}{\Lambda} L Y_B(\tau_B) N_B^c \right] H_u \nonumber\\
&&+ \left[ L Y_e(\tau_C) e^c + L Y_\mu(\tau_C) \mu^c + L Y_\tau(\tau_C) \tau^c \right] H_d \nonumber\\
&&+ \frac{1}{2} M_A(\tau_A) N_A^c N_A^c + \frac{1}{2} M_B(\tau_B) N_B^c N_B^c + M_{AB}(\tau_A,\tau_B) N_A^c N_B^c\,,
\end{eqnarray}
where terms such as  e.g., $L Y_A(\tau_A) N_A^c$ may be explicitly written as 
\begin{eqnarray}
L Y_A(\tau_A) N_A^c &=& \left[ L_1 (Y_A)_1 + L_2 (Y_A)_3 + L_3 (Y_A)_2 \right] N_A^c \,,
\end{eqnarray}
which is straightforwardly obtained from Eq.~\eqref{eq:example}.
This superpotential involves only the single residual $S_4^D$, and three modular fields $\tau_A$, $\tau_B$ and $\tau_C$ at the same time. 

The above superpotential may be taken as a starting point for 
models based on a single modular $S_4$ symmetry, where the 
three moduli fields introduced in an {\it ad hoc} way and taken to be independent fields.
However, we have shown that such a model can consistently arise from a high energy 
model involving three modular groups $S_4^3$.
The key point of such a model is that, in the low energy effective theory, 
the three moduli transform under the same $S_4^D$,  i.e., for any $\gamma_D \in S_4^D$, $\tau_A$, $\tau_B$ and $\tau_C$ transform in the following way,
\begin{eqnarray} \label{eq:modular_transformation_D}
\gamma_D: &&\tau_J \to \gamma_D \tau_J = \frac{a_D \tau_J + b_D}{c_D \tau_J + d_D}\,, 
\end{eqnarray}
for $J=A,B,C$. 
We also write out transformation properties of leptons 
\begin{eqnarray}
 L &\to& L(\gamma_D) = \rho_{\mathbf{3}}(\gamma_D) L \,,\nonumber\\
\alpha^c(\tau_C) &\to& \alpha^c(\gamma_D\tau_C) = (c_D \tau_C + d_D)^{-2k_\alpha} \alpha^c(\tau_C) \,,\nonumber\\
N_A^c(\tau_A) &\to& N_A^c(\gamma_D\tau_A) = (c_D \tau_A + d_D)^{-6} N_A^c(\tau_A) \,,\nonumber\\
N_B^c(\tau_B) &\to& N_B^c(\gamma_D\tau_B) = (c_D \tau_B + d_D)^{-4} N_B^c(\tau_B) \,,
  \label{eq:field_transformation_D}
\end{eqnarray}
and those for modular forms
\begin{eqnarray}
Y_\alpha(\tau_C) &\to& Y_\alpha(\gamma_D\tau_C) = (c_D \tau_C + d_D)^{2k_\alpha} \rho_{\mathbf{3}}(\gamma_D) Y_\alpha(\tau_C) \,,\nonumber\\
Y_A(\tau_A) &\to& Y_A(\gamma_D\tau_A) = (c_D \tau_A + d_D)^{6} \rho_{\mathbf{3}}(\gamma_D) Y_A(\tau_A) \,,\nonumber\\
Y_B(\tau_B) &\to& Y_B(\gamma_D\tau_B) = (c_D \tau_B + d_D)^{4} \rho_{\mathbf{3}}(\gamma_D) Y_B(\tau_B) \,,\nonumber\\
M_A(\tau_A) &\to& M_A(\gamma_D\tau_A) = (c_D \tau_A + d_D)^{12} M_A(\tau_A) \,,\nonumber\\
M_B(\tau_B) &\to& M_B(\gamma_D\tau_B) = (c_D \tau_B + d_D)^{8} M_B(\tau_B) \,,\nonumber\\
M_{AB}(\tau_A,\tau_B) &\to& M_{AB}(\gamma_D\tau_A, \gamma_D\tau_B) = (c_D \tau_A + d_D)^{6} (c_D \tau_B + d_D)^{4} M_{AB}(\tau_A,\tau_B) \,,  
 \label{eq:form_transformation_D}
\end{eqnarray}
where $\alpha=e,\mu,\tau$ and $k_{e,\mu,\tau} = 3,2,1$. 

We make a further comment on residual modular symmetries. It is well-known that in classical flavour model building, the residual symmetry for Majorana neutrinos is restricted to $Z_2$ or $Z_2 \times Z_2$. In the framework of modular symmetry, the residual  symmetry can be relaxed, e.g., $Z_3$ for $N_A$ as will be applied in section~\ref{sec:4.5}. And the reason is that the relevant mass is not a trivial coefficient but a modular form, which can vary with residual modular transformation. This novel feature could be applied to other phenomenological model constructions. For example, the residual symmetry to stabilise a dark matter candidate is not limited to a $Z_2$, while the latter is necessary in classic models of non-Abelian discrete symmetry \cite{Hirsch:2010ru}.

To summarise, we have derived a low energy effective flavon-less leptonic flavour model with one modular $S_4$ symmetry and three independent moduli fields. 
The importance of this for model building is that, as we shall see shortly,
by making use of the different moduli fields, we can access different sets of triplet modular forms,
corresponding to having different residual symmetries in different sectors of the theory.
This is similar to the traditional approach to model building based on $S_4$, but of course is achieved now without having to introduce flavons with certain vacuum alignments.

\subsection{Flavour structure in the charged lepton sector}

In the charged lepton sector, only $S_4^C$ plays a role. We assume the VEV of $\tau_C$ fixed at $\langle \tau_C \rangle = \tau_T =\omega$. Following Eq.~\eqref{eq:residual_T}, we obtain 
\begin{eqnarray}
Y_e(\langle\tau_C\rangle) =
\begin{pmatrix}
1\\
0\\
0
\end{pmatrix} \,,\quad
Y_\mu(\langle\tau_C\rangle) =
\begin{pmatrix}
0\\
0\\
1
\end{pmatrix}\,,\quad
Y_\tau(\langle\tau_C\rangle) =
\begin{pmatrix}
0\\
1\\
0
\end{pmatrix}\,,
\end{eqnarray}
for weights $2k_C=6, 4, 2$, respectively. This is a consequence of the residual modular  $Z_3^T$ symmetry. These modular forms will lead to diagonal Yukawa couplings for the charged leptons, where all lepton mixing arises from the neutrino sector.
Although the diagonal Yukawa couplings are independent, we do not gain any understanding of the charged lepton
mass hierarchy in this model.

\subsection{Flavour structure in the neutrino sector \label{sec:4.5}}

In the neutrino sector, by selecting 
$\langle \tau_A \rangle = \tau_{TS} = \frac{1}{2}+i\frac{\sqrt{3}}{2}$ and $\langle \tau_B \rangle = \tau_{U} = \frac{1}{2}+\frac{i}{2}$ we have residual modular symmetries $Z_3^{TS}$ and $Z_2^U$, respectively. Following the discussion in section~\ref{sec:residual}, we obtain the modular form for the Yukawa coupling
\begin{eqnarray}
Y_A(\langle\tau_A\rangle) =
\begin{pmatrix}
-1\\
2 \omega\\
2 \omega^2
\end{pmatrix} \,,\quad
Y_B(\langle\tau_B\rangle) =
\begin{pmatrix}
0\\
1\\
-1
\end{pmatrix}\,,
\end{eqnarray}
by selecting the modular weights of $N_A^c$ and $N_B^c$ in $S_4^A$ and $S_4^B$ to be $2k_A = -6$ and $2k_B = -4$, respectively. $Y_A(\langle\tau_A\rangle)$ and $Y_B(\langle\tau_B\rangle)$ give rise to the $3\times 2$ Dirac neutrino mass matrix $M_D'$. $M_A$, $M_B$ and $M_{AB}$ all takes non-zero values at $\langle \tau_A \rangle = \tau_{TS}$ and $\langle \tau_B \rangle = \tau_{U}$. Thus, we obtain a $2 \times 2$ Majorana matrix for $N_A^c$ and $N_B^c$,
\begin{eqnarray} 
M_N = \begin{pmatrix} M_A & M_{AB} \\ M_{AB} & M_B \end{pmatrix} \,.
\end{eqnarray}
Here, we still use $M_{A}$, $M_{B}$ and $M_{AB}$ to represent values of $M_{A}(\tau_A)$, $M_{B}(\tau_B)$ and $M_{AB}(\tau_A,\tau_B)$ at the relevant VEVs. $M_N$ can be diagonalised by a unitary matrix $V$ via $V^T M_N V = {\rm diag}\{M_1, M_2\}$, with 
\begin{eqnarray} 
V = e^{i \alpha_3}
\begin{pmatrix} \hat{c}_R & \hat{s}_R^* \\ -\hat{s}_R & \hat{c}_R^* \end{pmatrix} \,,
\end{eqnarray}
where $\hat{c}_R \equiv \cos \theta_R e^{\alpha_1}$ and $\hat{s}_R \equiv \sin \theta_R e^{i\alpha_2}$.
The Dirac mass matrix $M_D$ in the basis where charged lepton and right-handed neutrino mass matrices are diagonal is obtained through $V$ acting on the right of $M_D'$, which mixes the columns:
\begin{eqnarray}
M_D = e^{i \alpha_3}
\begin{pmatrix} 
-\hat{c}_R & - \hat{s}_R^* \\
2\omega^2 \hat{c}_R + \hat{s}_R~ & ~2\omega^2 \hat{s}_R^* - \hat{c}_R^* \\
2\omega \hat{c}_R - \hat{s}_R~ & ~2\omega \hat{s}_R^* + \hat{c}_R^* 
\end{pmatrix}\,.
\end{eqnarray}
Applying seesaw formula, we obtain
\begin{eqnarray}
M_\nu &=& (\mu_1 \hat{c}_R^2 + \mu_2 \hat{s}_R^{*2}) 
\begin{pmatrix}
 1 & -2 \omega ^2 & -2 \omega  \\
 -2 \omega ^2 & 4 \omega  & 4 \\
 -2 \omega  & 4 & 4 \omega ^2 \\
\end{pmatrix}
+(\mu_1 \hat{s}_R^2 + \mu_2 \hat{c}_R^{*2}) 
\begin{pmatrix}
 0 & 0 & 0 \\
 0 & 1 & -1 \\
 0 & -1 & 1 \\
\end{pmatrix} \nonumber\\
&& + (\mu_1 \hat{c}_R\hat{s}_R - \mu_2 \hat{c}_R^{*} \hat{s}_R^{*} )
\begin{pmatrix}
 0 & -1 & 1 \\
 -1 & 4 \omega ^2 & 2 i \sqrt{3} \\
 1 & 2 i \sqrt{3} & -4 \omega  \\
\end{pmatrix}
 \,,
\end{eqnarray}
where $\mu_1$ and $\mu_2$ are real inspect of an overall phase. There are five physical parameters $\mu_1$, $\mu_2$, $\theta_R$, $\alpha_1$ and $\alpha_2$. 

The PMNS matrix is obtained by diagonalising the neutrino mass matrix, $U^T M_\nu U = {\rm diag} \{ 0, m_2, m_3 \}$. 
Since both $Y_A$ and $Y_B$ are orthogonal to $(2,-1,-1)^T$, we directly arrive at the TM$_1$ 
form of lepton
mixing matrix~\cite{Xing:2006ms, Lam:2006wm, Albright:2008rp, Albright:2010ap},
\begin{equation}\label{TMM}
\!\!\!\!\!\!\!\!
U_{\rm  TM_1} =
\left(
\begin{array}{ccc}
\frac{2}{\sqrt{6}} &  - &  - \\ 
-\frac{1}{\sqrt{6}} &  - &  - \\
-\frac{1}{\sqrt{6}} &  - &  -  
\end{array}
\right).
\end{equation}%
$\rm{TM}_1$ lepton
mixing implies three equivalent relations:
\begin{equation}
\tan \theta_{12} = \frac{1}{\sqrt{2}}\sqrt{1-3s^2_{13}}\ \ \ \ {\rm or} \ \ \ \ 
\sin \theta_{12}= \frac{1}{\sqrt{3}}\frac{\sqrt{1-3s^2_{13}}}{c_{13}} \ \ \ \ {\rm or} \ \ \ \ 
\cos \theta_{12}= \sqrt{\frac{2}{3}}\frac{1}{c_{13}}
\label{t12p}
\end{equation}
leading to a prediction $\theta_{12}\approx 34^{\circ}$,
in excellent agreement with current global fits, assuming $\theta_{13}\approx 8.5^{\circ}$.
By contrast, the corresponding $\rm{TM}_2$ relations imply $\theta_{12}\approx 36^{\circ}$ \cite{Albright:2008rp}, which is on the edge of the three sigma region, and hence disfavoured by current data.
$\rm{TM}_1$ mixing also leads to an exact sum rule relation relation for $\cos \delta$ in terms of the other lepton mixing angles
\cite{Albright:2008rp},
\begin{equation}
\cos \delta = - \frac{\cot 2\theta_{23}(1-5s^2_{13})}{2\sqrt{2}s_{13}\sqrt{1-3s^2_{13}}} \,.
\label{TM1sum}
\end{equation}

\subsection{Numerical fit}

As described in previous subsections, we obtain through the use of modular symmetries a flavon-less effective theory which fulfils TM$_1$ lepton mixing. In this section, we make use of the above analytical sum rules for TM$_1$ lepton mixing as well as the diagonalisation of the $2 \times 2$ symmetric matrices which result from the rotation of the neutrino mass matrix by the 
TB mixing matrix, following the analytic methods presented in~\cite{King:2015dvf}. We are thus able to express each observable (the 3 mixing angles, the squared mass ratio and the CP-violating phase $\delta$) in terms of the model parameters $(\{ x \}) = (\{\alpha_1, \alpha_2, \theta_R, \mu_1, \mu_2 \})$, i.e. the phases $\alpha_1$ and $\alpha_2$, the angle parametrizing the rotation originating from RH neutrino sector, and the parameters governing the contribution from $Y_A$ and $Y_B$, $\mu_1$ and $\mu_2$. These formulas are somewhat complicated and not particularly illustrative, but enable us to easily run a numerical minimisation procedure on a $\chi^2$ function:
\begin{equation}
\label{eq:chi2}
\chi^2 = \sum \left( \frac{P_i(\{ x \}) - {\rm BF}_i}{\sigma_i} \right)^2 \,,
\end{equation}
where $P_i$ are the model predictions, BF$_i$ the current best-fit values, and the errors $\sigma_i$ correspond here to the average of the $1 \sigma$ ranges for each observable. We use the best-fit values and $1 \sigma$ ranges from NuFit 4.0 \cite{Esteban:2018azc, nufit4}.
The minimisation runs over model parameters and the observables tested are the 3 PMNS mixing angles, the phase $\delta$, and the absolute masses obtained from the square roots of the squared mass differences (taking into account that we have only 2 RH neutrinos and normal mass ordering, $m_1 = 0$).

The obtained best-fit point (BF) corresponds to a $\chi^2 = 0.74$, with the model parameters shown in Table \ref{ta:benchmarks}, together with the respective predictions for the observables, including mixing parameters, neutrino masses, and the effective neutrino mass parameter in neutrino-less double beta decay $m_{ee}= |\mu_1 \hat{c}_R^2 + \mu_2 \hat{s}_R^{*2}|$. These observables (predicted by the analytical formulas for the specific point in parameter space) completely match with the values obtained by performing an entirely numerical diagonalisation for the same point in parameter space.
For the best-fit point the observables are all within the $1 \sigma$ range except $\delta = 290^\circ$.

For comparison we present also two other benchmark points. In Benchmark 1 (B1), the observables are all within the $1 \sigma$ range except $\delta$, which is slightly smaller ($285^\circ$) than in the best fit point. Conversely, $\theta_{23}$ deviates slightly from its best-fit point. The total $\chi^2 = 1.6$ is slightly worse.
In Benchmark 2  (B2), $\delta = 254^\circ$ is within the $1 \sigma$ range. Conversely, $\theta_{13}$ is slightly deviated from its best fit point and $\theta_{23} = 41.5^\circ$ is strongly deviated from its best-fit point and is indeed outside the $1 \sigma$ range. The total $\chi^2=55$ is much worse, although we note that this value is somewhat spurious, given that the expression in Eq.~\eqref{eq:chi2} is based on Gaussian distributions, which is not the case for $\theta_{23}$, which for B2 contributes 0.99 of the total $\chi^2$. We are taking the best-fit point for $\theta_{23} = 49.6^\circ$ from NuFit 4.0 \cite{Esteban:2018azc, nufit4}.

It is worth emphasizing that these predictions originate from the special directions $Y_A$ and $Y_B$ obtained from the fixed points in the respective modular symmetries. The best-fit point observables all lie within the $1 \sigma$ range except $\delta$, which nevertheless lies within its $3 \sigma$ range and takes a value close to maximal ($290^\circ$).

\begin{table}
\centering
\begin{tabular}{| c|c | c |}
\hline \hline
\multirow{3}{*}{BF} & Para.
& 
\begin{tabular}{| c | ccccc}
$\chi^2$  & $\alpha_1$ & $\alpha_2$ & $\theta_R$ & $\mu_1$ & $\mu_2$ \\ \hline
0.74 & $64.53^\circ$ & $20.38^\circ$ & $43.01^\circ$ & $0.00633$\,eV & $0.0114$\,eV \\ 
\end{tabular}
\\\cline{2-3}
& Obs. 
&
\begin{tabular}{ccccccc}
$\theta_{12}$ & $\theta_{13}$ & $\theta_{23}$ & $\delta$ & $m_2$ & $m_3$ & $m_{ee}$ \\  \hline
$34.33^\circ$ & $8.61^\circ$ & $49.6^\circ$ & $290^\circ$ &  $0.00860$\,eV & $0.0502$\,eV & $0.00206$\,eV
\end{tabular}
\\
\hline \hline
\multirow{3}{*}{B1} & Para.
& 
\begin{tabular}{| c | ccccc}
$\chi^2$ & $\alpha_1$ & $\alpha_2$ & $\theta_R$ & $\mu_1$ & $\mu_2$ \\ \hline
1.6 & $70.16^\circ$ & $16.62^\circ$ & $43.51^\circ$ & $0.00651$\,eV & $0.0135$\,eV \\ 
\end{tabular}
\\\cline{2-3}
& Obs.
&
\begin{tabular}{ccccccc}
$\theta_{12}$ & $\theta_{13}$ & $\theta_{23}$ & $\delta$ & $m_2$ & $m_3$ & $m_{ee}$ \\  \hline
$34.33^\circ$ & $8.62^\circ$ & $48.6^\circ$ & $285^\circ$ &  $0.00860$\,eV & $0.0502$\,eV & $0.00188$\,eV
\end{tabular}
\\
\hline \hline
\multirow{3}{*}{B2} & Para.
& 
\begin{tabular}{| c | ccccc}
$\chi^2$ & $\alpha_1$ & $\alpha_2$ & $\theta_R$ & $\mu_1$ & $\mu_2$ \\ \hline
55 & $358.73^\circ$ & $338.89^\circ$ & $24.65^\circ$ & $0.00533$\,eV & $0.0114$\,eV \\ 
\end{tabular}
\\\cline{2-3}
& Obs. 
&
\begin{tabular}{ccccccc}
$\theta_{12}$ & $\theta_{13}$ & $\theta_{23}$ & $\delta$ & $m_2$ & $m_3$ & $m_{ee}$ \\  \hline
$34.34^\circ$ & $8.56^\circ$ & $41.5^\circ$ & $254^\circ$ &  $0.00860$\,eV & $0.0502$\,eV & $0.00319$\,eV
\end{tabular}
\\
\hline \hline

\end{tabular}
\caption{Model parameters (Para.) and respective observables (Obs.) for the best-fit point (BF) and two other benchmark points (B1, B2). \label{ta:benchmarks}}
\end{table}

\section{Conclusions and Discussion}
\label{conclusion}

In this paper we have considered, for the first time, leptonic flavour models based on multiple moduli fields
with an extended finite modular symmetry. 
We reviewed the case of
a single modular symmetry $\overline{\Gamma}$
with a single modulus field $\tau$ and $\mathcal{N} = 1$ supersymmetry,
then extended the formalism to include 
a series of modular groups $\overline{\Gamma}^{1}$, $\overline{\Gamma}^{2}$, ..., $\overline{\Gamma}^{M}$, where the modulus field for each modular symmetry $\overline{\Gamma}^{J}$ is denoted as $\tau_J$,
where $J=1,..., M$, 
resulting in the finite modular symmetry
$\Gamma_{N_1}^1\times \Gamma_{N_2}^2 \times \cdots \times \Gamma_{N_M}^M$.

We then returned to the case of a single modular symmetry, focussing on
the case of modular $S_4$ symmetry and its remnant symmetries, exploring relations of stabilisers of modular transformations, residual symmetries and modular forms in the framework of finite modular symmetry. 
In the case of modular $S_4$ symmetry, several new stabilisers of residual symmetries were identified,
where each stabiliser preserves a $Z_2$ or $Z_3$ residual symmetry. 
We discovered a strong correlation between the modular transformation and the modular form at its stabiliser,
namely that 
{\it a modular form at a stabiliser of any modular transformation is an eigenvector of the representation matrix of the modular transformation.} Based on this correlation, we were able to determine some new types of modular forms without knowing exact expressions for those modular forms. 

As an application of the preceding results, we constructed a flavour model of leptons involving 
two right-handed neutrinos and three finite modular symmetries $S_4^A \times S_4^B \times S_4^C$. Here, $S_4^A$ and $S_4^B$ are modular symmetries for two right-handed neutrinos, respectively, while $S_4^C$ is the modular symmetry in charged lepton sector. They are connected by two bi-triplet scalars. 
After they gain VEVs, three $S_4$'s are broken to a single $S_4^D$, i.e., $S_4^A \times S_4^B \times S_4^C \to S_4^D$. 
Independent fixed points in the extra dimensions associated with $S_4^A$ and $S_4^B$ specify (flavon-less) special directions that preserve subgroups of the respective symmetries, whereas a scalar transforming as a triplet of both $S_4^A$ and $S_4^C$ and another scalar transforming one of both as a triplet of both $S_4^B$ and $S_4^C$ acquire vacuum expectation values that break $S_4^A \times S_4^B \times S_4^C$ to its diagonal subgroup $S_4^D$.
We emphasise that these scalars do not carry any information about flavour.

After the three $S_4$'s are broken, we arrive at an effective low energy 
flavour mixing model with a single $S_4$ modular symmetry but three independent modular fields $\tau_A$, $\tau_B$ and $\tau_C$. The independence of these modular fields allows us to assign different VEVs for them
which determine the flavour structure. We fix the VEV of $\tau_C$ at a stabiliser which satisfies a modular $Z^C_3$ symmetry. A diagonal charged lepton mass matrix is obtained. VEVs of $\tau_A$ and $\tau_B$ are fixed at other two stabilisers which preserve a different $Z^A_3$ symmetry and a $Z^B_2$ symmetry respectively. 
The residual modular symmetries justify the special directions that lead to TM$_1$ mixing. 
This is similar to the traditional approach to model building based on $S_4$, but of course is achieved now without having to introduce flavons with certain vacuum alignments.

Finally, we performed an analysis of the predictions of the model taking into account the existence of RH neutrino mixing (in the model-building basis). When this is taken into account, the 5 observables depend on 4 real model parameters, and we obtain an excellent fit to experiment, with all 3 mixing angles and the squared mass ratio within $1 \sigma$ of their experimental values and a near-maximal value for $\delta = 290$ degrees. Having two right-handed neutrinos, the model predicts the absolute neutrino mass scale $m_1 = 0$.

In conclusion, we have developed a general formalism for multiple modular symmetries, analysed the residual symmetries
of modular $S_4$ symmetry, and proposed a realistic model based on modular $S_4^3$ symmetry,
which yields the successful trimaximal TM$_1$ lepton mixing, without requiring any flavons.

%%%%%%%%%%%%%%%%%%%%%%%%%%%%%%%%%%%%%%%%%%
%%%%%%%%%%%%%%%%%%%%%%%%%%%%%%%%%%%%%%%%%%

\subsection*{Acknowledgements}
IdMV acknowledges funding from the Funda\c{c}\~{a}o para a Ci\^{e}ncia e a Tecnologia (FCT) through the contract IF/00816/2015 and partial support by FCT through projects CFTP-FCT Unit 777 (UID/FIS/00777/2019), CERN/FIS-PAR/0004/2017 and PTDC/FIS-PAR/29436/2017 which are partially funded through POCTI (FEDER), COMPETE, QREN and EU. SFK and YLZ acknowledge the STFC Consolidated Grant ST/L000296/1 and the European Union's Horizon 2020 Research and Innovation programme under Marie Sk\l{}odowska-Curie grant agreements Elusives ITN No.\ 674896 and InvisiblesPlus RISE No.\ 690575.

%%%%%%%%%%%%%%%%%%%%%%%%%%%%%%%%%%%%%%%%%%
%%%%%%%%%%%%%%%%%%%%%%%%%%%%%%%%%%%%%%%%%%

\appendix

\section{Group theory of $S_4$ \label{app:S4} } 

$S_4$ is the permutation group of 4 objects, see e.g.\ \cite{Escobar:2008vc}.
The Kronecker products between different irreducible representations can be easily obtained:
\begin{eqnarray}
&\mathbf{1^{\prime}}\otimes\mathbf{1^{\prime}}=\mathbf{1}, ~~\mathbf{1^{\prime}}\otimes\mathbf{2}=\mathbf{2}, ~~\mathbf{1^{\prime}}\otimes\mathbf{3}=\mathbf{3^{\prime}},~~ 
\mathbf{1^{\prime}}\otimes\mathbf{3^{\prime}}=\mathbf{3},\nonumber\\
&\mathbf{2}\otimes\mathbf{2}=\mathbf{1}\oplus\mathbf{1}^{\prime}\oplus\mathbf{2},~~
\mathbf{2}\otimes\mathbf{3}=\mathbf{2}\otimes\mathbf{3^{\prime}}=\mathbf{3}\oplus\mathbf{3}^{\prime},\nonumber\\
&\mathbf{3}\otimes\mathbf{3}=\mathbf{3^{\prime}}\otimes\mathbf{3^{\prime}}=\mathbf{1}\oplus \mathbf{2}\oplus\mathbf{3}\oplus\mathbf{3^{\prime}},~~
\mathbf{3}\otimes\mathbf{3^{\prime}}=\mathbf{1^{\prime}}\oplus \mathbf{2}\oplus\mathbf{3}\oplus\mathbf{3^{\prime}}\,.
\end{eqnarray}
%
%%%%%%%%%%%%%%%
\begin{table}[h!]
\begin{center}
\begin{tabular}{|c|ccc|}
\hline\hline
   & $\rho(T)$ & $\rho(S)$ & $\rho(U)$  \\\hline
$\mathbf{1}$ & 1 & 1 & 1 \\
$\mathbf{1^{\prime}}$ & 1 & 1 & $-1$ \\
$\mathbf{2}$ & 
$\left(
\begin{array}{cc}
 \omega  & 0 \\
 0 & \omega ^2 \\
\end{array}
\right)$ & 
$\left(
\begin{array}{cc}
 1 & 0 \\
 0 & 1 \\
\end{array}
\right)$ & 
$\left(
\begin{array}{cc}
 0 & 1 \\
 1 & 0 \\
\end{array}
\right)$ \\

$\mathbf{3}$ &  $\left(
\begin{array}{ccc}
 1 & 0 & 0 \\
 0 & \omega ^2 & 0 \\
 0 & 0 & \omega  \\
\end{array}
\right)$ &
$\frac{1}{3} \left(
\begin{array}{ccc}
 -1 & 2 & 2 \\
 2 & -1 & 2 \\
 2 & 2 & -1 \\
\end{array}
\right)$ &
$\left(
\begin{array}{ccc}
 1 & 0 & 0 \\
 0 & 0 & 1 \\
 0 & 1 & 0 \\
\end{array}
\right)$ \\

$\mathbf{3^{\prime}}$ &  $\left(
\begin{array}{ccc}
 1 & 0 & 0 \\
 0 & \omega ^2 & 0 \\
 0 & 0 & \omega  \\
\end{array}
\right)$ &
$\frac{1}{3} \left(
\begin{array}{ccc}
 -1 & 2 & 2 \\
 2 & -1 & 2 \\
 2 & 2 & -1 \\
\end{array}
\right)$ &
$-\left(
\begin{array}{ccc}
 1 & 0 & 0 \\
 0 & 0 & 1 \\
 0 & 1 & 0 \\
\end{array}
\right)$ \\ \hline\hline

\end{tabular}
\caption{\label{tab:rep_matrix_main} The representation matrices for the $S_4$ generators $T$, $S$ and $U$ used in the main text, where $\omega$ is the cube root of unity $\omega=e^{2\pi i/3}$.}
\end{center}
\end{table}
%%%%%%%%%%%%%%%%%%%%%%%%
% 

The generators of $S_4$ in different irreducible representations are listed in Table \ref{tab:rep_matrix_main}, in the basis we used in the main text.  
This basis is widely used in the literature since the charged lepton mass matrix invariant under $T$ is diagonal in this basis. The following basis-dependent property is satisfied, $\rho^T_{\mathbf{3}^{(\prime)}}(\gamma) P_{23} = P_{23} \rho_{\mathbf{3}^{(\prime)}}(\gamma^{-1})$ for any $\gamma$ of $S_4$.  The products of two 3 dimensional irreducible representations $a$ and $b$ can be expressed as
\begin{eqnarray}
(ab)_\mathbf{1_i} &=& a_1b_1 + a_2b_3 + a_3b_2 \,,\nonumber\\
(ab)_\mathbf{2} &=& (a_2b_2 + a_1b_3 + a_3b_1,~ a_3b_3 + a_1b_2 + a_2b_1)^T \,,\nonumber\\
(ab)_{\mathbf{3_i}} &=& (2a_1b_1-a_2b_3-a_3b_2, 2a_3b_3-a_1b_2-a_2b_1, 2a_2b_2-a_3b_1-a_1b_3)^T \,,\nonumber\\
(ab)_{\mathbf{3_j}} &=& (a_2b_3-a_3b_2, a_1b_2-a_2b_1, a_3b_1-a_1b_3)^T \,,
\label{eq:CG2}
\end{eqnarray}
where 
\begin{eqnarray} \label{eq:_i_j}
&&\mathbf{1_i}=\mathbf{1}\,, ~\; \mathbf{3_i}=\mathbf{3}\,, ~\; \mathbf{3_j}=\mathbf{3'}\,~\; \text{for} ~\; a\sim b \sim \mathbf{3}\,,~ \mathbf{3^\prime} \,, \nonumber\\
&&\mathbf{1_i}=\mathbf{1'}\,,~  \mathbf{3_i}=\mathbf{3'}\,,~ \mathbf{3_j}=\mathbf{3}\,~\;\; \text{for} ~\; a\sim  \mathbf{3}\,,~ b \sim \mathbf{3'}\,.
\end{eqnarray} 

The products of two doublets $a=(a_1, a_2)^T$ and $b=(b_1, b_2)^T$ are divided into
\begin{eqnarray}
(ab)_\mathbf{1} &=& a_1b_2 + a_2b_1 \,,\quad
(ab)_\mathbf{1^\prime} = a_1b_2 - a_2b_1 \,,\quad
(ab)_{\mathbf{2}} = (a_2b_2, a_1b_1)^T \,.
\label{eq:CG_doublets}
\end{eqnarray}

%%%%%%%%%%%%%%%
\begin{table}[h!]
\begin{center}
\begin{tabular}{|c|ccc|}
\hline\hline
   & $\tilde{\rho}(T)$ & $\tilde{\rho}(S)$ & $\tilde{\rho}(U)$  \\\hline
$\mathbf{1}$ & 1 & 1 & 1 \\
$\mathbf{1^{\prime}}$ & 1 & 1 & $-1$ \\
$\mathbf{2}$ & 
$\left(
\begin{array}{cc}
 \omega  & 0 \\
 0 & \omega ^2 \\
\end{array}
\right)$ & 
$\left(
\begin{array}{cc}
 1 & 0 \\
 0 & 1 \\
\end{array}
\right)$ & 
$\left(
\begin{array}{cc}
 0 & 1 \\
 1 & 0 \\
\end{array}
\right)$ \\

$\mathbf{3}$ &  $\left(
\begin{array}{ccc}
 0 & 0 & 1 \\
 1 & 0 & 0 \\
 0 & 1 & 0 \\
\end{array}
\right)$ & 
$\left(
\begin{array}{ccc}
 1 & 0 & 0 \\
 0 & -1 & 0 \\
 0 & 0 & -1 \\
\end{array}
\right)$ &
$\left(
\begin{array}{ccc}
 1 & 0 & 0 \\
 0 & 0 & 1 \\
 0 & 1 & 0 \\
\end{array}
\right)$

\\

$\mathbf{3^{\prime}}$ &  
$\left(
\begin{array}{ccc}
 0 & 0 & 1 \\
 1 & 0 & 0 \\
 0 & 1 & 0 \\
\end{array}
\right)$ & 
$\left(
\begin{array}{ccc}
 1 & 0 & 0 \\
 0 & -1 & 0 \\
 0 & 0 & -1 \\
\end{array}
\right)$ &
$-\left(
\begin{array}{ccc}
 1 & 0 & 0 \\
 0 & 0 & 1 \\
 0 & 1 & 0 \\
\end{array}
\right)$  \\ \hline\hline

\end{tabular}
\caption{\label{tab:rep_matrix_vacuum} The representation matrices for the $S_4$ generators $T$, $S$ and $U$ used for vacuum alignments.}
\end{center}
\end{table}
%%%%%%%%%%%%%%%%%%%%%%%%

In Appendix~\ref{eq:vacuum}, we apply another basis to calculate the vacuum alignment. Both bases are widely used and have no physical difference. However, we apply this basis because it is simpler to carry out the respective calculations. Representation matrices for generators in this basis are listed in Table~\ref{tab:rep_matrix_vacuum}. Representation matrix for any $\gamma$ of $S_4$ satisfies $\tilde{\rho}^T_{\mathbf{3}^{(\prime)}}(\gamma) = \tilde{\rho}_{\mathbf{3}^{(\prime)}}(\gamma^{-1})$. 
Basis transformation between the first and second basis are given by
\begin{eqnarray} \label{eq:basis_transformation}
\tilde{\rho}_{\mathbf{3}^{(\prime)}}(\gamma)  = U_\omega \rho_{\mathbf{3}^{(\prime)}}(\gamma) U_\omega^{\dag}
\end{eqnarray}
for any $\gamma \in S_4$, where
\begin{eqnarray}
U_\omega = \frac{1}{\sqrt{3}}\begin{pmatrix} 
 1 & 1 & 1 \\
 1 & \omega ^2 & \omega  \\
 1 & \omega  & \omega ^2 \\
\end{pmatrix} \,.
\end{eqnarray}
Irreducible products of two triplet representations $\tilde{a}$ and $\tilde{b}$ are simply written as 
\begin{eqnarray}
(\tilde{a}\tilde{b})_\mathbf{1_i} &=& \sum_{i=1,2,3} \tilde{a}_i \tilde{b}_i \,, \nonumber\\
\left((\tilde{a}\tilde{b})_\mathbf{2}\right)_{i} &=& \sum_{j=1,2,3} \omega^{i(j-1)} \tilde{a}_i \tilde{b}_i ~\text{ for } i=1,2 \,, \nonumber\\
\left((\tilde{a}\tilde{b})_\mathbf{3_i}\right)_{i} &=& \sum_{j,k=1,2,3}|\epsilon_{ijk}|\tilde{a}_j \tilde{b}_k ~\text{ for } i=1,2,3\,, \nonumber\\
\left((\tilde{a}\tilde{b})_{\mathbf{3_j}}\right)_{i} &=& \sum_{j,k=1,2,3}\epsilon_{ijk}\tilde{a}_j \tilde{b}_k ~\text{ for } i=1,2,3\,, 
\label{eq:CG_vacuum}
\end{eqnarray}
where $\mathbf{1_i}$, $\mathbf{3_i}$ and $\mathbf{3_j}$ are given the same as in Eq.~\eqref{eq:_i_j}.

%%%%%%%%%%%%%%%%%%%%%%%%

\section{Vacuum alignments \label{eq:vacuum}}

\begin{table}
\centering
\begin{tabular}{| l | c c c c c c|}
\hline \hline
Fields & $S_4^A$ & $S_4^B$ & $S_4^C$ & $2k_A$ & $2k_B$ & $2k_C$\\ 
\hline \hline
$\chi_{AC}$ & $\mathbf{3}$ & $\mathbf{1}$ & $\mathbf{3}$ & 0 & 0 & 0 \\
$\chi_{BC}$ & $\mathbf{1}$ & $\mathbf{3}$ & $\mathbf{3}$ & 0 & 0 & 0 \\
$\chi_{A}$ & $\mathbf{3}$ & $\mathbf{1}$ & $\mathbf{1}$ & 0 & 0 & 0 \\
$\chi_{B}$ & $\mathbf{1}$ & $\mathbf{3}$ & $\mathbf{1}$ & 0 & 0 & 0
\\
\hline \hline
\end{tabular}
\caption{Symmetry and field content according to sector (scalar and driving sectors), for fields responsible for vacuum alignment.}
\label{tab:scalars}
\end{table}

Vacuum alignments for the bi-triplet scalars $\Phi_{AC}$ and $\Phi_{BC}$ can be realised following the general way in most supersymmetric flavour models. 
As seen in Table~\ref{tab:scalars}, we introduce four driving fields $\chi_{AC}$, $\chi_{A}$ and $\chi_{BC}$, $\chi_{B}$, 
\begin{eqnarray}
\chi_{AC} \sim (\mathbf{3}, \mathbf{1}, \mathbf{3})\,, && \chi_{A} \sim (\mathbf{3}, \mathbf{1}, \mathbf{1})\,, \nonumber\\
\chi_{BC} \sim (\mathbf{1}, \mathbf{3}, \mathbf{3})\,, && \chi_{B} \sim (\mathbf{1}, \mathbf{3}, \mathbf{1}) 
\end{eqnarray}
of $S_4^A \times S_4^B \times S_4^C$. The superpotential for vacuum alignment is given by
\begin{eqnarray}
w_d &=& \Phi_{AC}\Phi_{AC} \chi_{AC} + {\rm M}_A \Phi_{AC} \chi_{AC} + \Phi_{AC}\Phi_{AC} \chi_A \,, \nonumber\\
&+& \Phi_{BC}\Phi_{BC} \chi_{BC} + {\rm M}_B \Phi_{BC} \chi_{BC} + \Phi_{BC} \Phi_{BC} \chi_B \,,
\end{eqnarray}
where ${\rm M}_A$ and ${\rm M}_B$ are mass-dimensional coefficients. 
Minimisation of the superpotential gives rise to conditions for $\Phi_{AC}$ and $\Phi_{BC}$ VEVs. 
In order to give a better illustration, we present the derivation of $\Phi_{AC}$ in the second basis as $(\tilde{\Phi}_{AC})_{i\alpha} = \sum_{j,\beta} (U_\omega)_{ij} (U_\omega)_{\alpha\beta} \, (\Phi_{AC})_{j\beta}$. 
With the help of the Clebsch-Gordan coefficients in Eq.~\eqref{eq:CG_vacuum}, these conditions are explicitly written as 
\begin{eqnarray}
&&\sum_{j,k=1,2,3;} \sum_{\beta,\gamma=1,2,3} |\epsilon_{ijk}| |\epsilon_{\alpha\beta\gamma}| (\tilde{\Phi}_{AC})_{j \beta} (\tilde{\Phi}_{AC})_{k c} + {\rm M}_A  (\tilde{\Phi}_{AC})_{i \alpha} = 0  ~\text{ for } i=1,2,3,~ \alpha=1,2,3\,,\nonumber\\
&&\sum_{j,k=1,2,3;} \sum_{\alpha=1,2,3} |\epsilon_{ijk}| (\tilde{\Phi}_{AC})_{j \alpha} (\tilde{\Phi}_{AC})_{k \alpha} = 0 ~\text{ for } i=1,2,3 \,.
\end{eqnarray}
The full solution for the above equation is not hard to obtained. There are 24 solutions in total. It is convenient to write them as $3\times3$ unitary matrices,  
\begin{eqnarray} \label{eq:vacuum_all}
\langle \tilde{\Phi}_{AC} \rangle & = &\left\{
\left(
\begin{array}{ccc}
 1 & 0 & 0 \\
 0 & 1 & 0 \\
 0 & 0 & 1 \\
\end{array}
\right),\left(
\begin{array}{ccc}
 1 & 0 & 0 \\
 0 & -1 & 0 \\
 0 & 0 & -1 \\
\end{array}
\right),\left(
\begin{array}{ccc}
 -1 & 0 & 0 \\
 0 & -1 & 0 \\
 0 & 0 & 1 \\
\end{array}
\right),\left(
\begin{array}{ccc}
 -1 & 0 & 0 \\
 0 & 1 & 0 \\
 0 & 0 & -1 \\
\end{array}
\right),\right. \nonumber\\ &&
\left(
\begin{array}{ccc}
 1 & 0 & 0 \\
 0 & 0 & -1 \\
 0 & -1 & 0 \\
\end{array}
\right),\left(
\begin{array}{ccc}
 1 & 0 & 0 \\
 0 & 0 & 1 \\
 0 & 1 & 0 \\
\end{array}
\right),
\left(
\begin{array}{ccc}
 0 & 0 & -1 \\
 0 & -1 & 0 \\
 1 & 0 & 0 \\
\end{array}
\right),\left(
\begin{array}{ccc}
 0 & 0 & -1 \\
 0 & 1 & 0 \\
 -1 & 0 & 0 \\
\end{array}
\right),\nonumber\\ &&
\left(
\begin{array}{ccc}
 0 & 0 & -1 \\
 -1 & 0 & 0 \\
 0 & 1 & 0 \\
\end{array}
\right),\left(
\begin{array}{ccc}
 0 & 0 & -1 \\
 1 & 0 & 0 \\
 0 & -1 & 0 \\
\end{array}
\right),\left(
\begin{array}{ccc}
 0 & 0 & 1 \\
 0 & -1 & 0 \\
 -1 & 0 & 0 \\
\end{array}
\right),\left(
\begin{array}{ccc}
 0 & 0 & 1 \\
 0 & 1 & 0 \\
 1 & 0 & 0 \\
\end{array}
\right)\,,\nonumber\\ &&
\left(
\begin{array}{ccc}
 0 & 0 & 1 \\
 -1 & 0 & 0 \\
 0 & -1 & 0 \\
\end{array}
\right),\left(
\begin{array}{ccc}
 0 & 0 & 1 \\
 1 & 0 & 0 \\
 0 & 1 & 0 \\
\end{array}
\right),\left(
\begin{array}{ccc}
 0 & -1 & 0 \\
 0 & 0 & -1 \\
 1 & 0 & 0 \\
\end{array}
\right),\left(
\begin{array}{ccc}
 0 & -1 & 0 \\
 0 & 0 & 1 \\
 -1 & 0 & 0 \\
\end{array}
\right),\nonumber\\ &&
\left(
\begin{array}{ccc}
 0 & -1 & 0 \\
 -1 & 0 & 0 \\
 0 & 0 & 1 \\
\end{array}
\right),\left(
\begin{array}{ccc}
 0 & -1 & 0 \\
 1 & 0 & 0 \\
 0 & 0 & -1 \\
\end{array}
\right),\left(
\begin{array}{ccc}
 0 & 1 & 0 \\
 0 & 0 & -1 \\
 -1 & 0 & 0 \\
\end{array}
\right),\left(
\begin{array}{ccc}
 0 & 1 & 0 \\
 0 & 0 & 1 \\
 1 & 0 & 0 \\
\end{array}
\right),\nonumber\\ &&
\left.\left(
\begin{array}{ccc}
 0 & 1 & 0 \\
 -1 & 0 & 0 \\
 0 & 0 & -1 \\
\end{array}
\right),\left(
\begin{array}{ccc}
 0 & 1 & 0 \\
 1 & 0 & 0 \\
 0 & 0 & 1 \\
\end{array}
\right),\left(
\begin{array}{ccc}
 -1 & 0 & 0 \\
 0 & 0 & -1 \\
 0 & 1 & 0 \\
\end{array}
\right),\left(
\begin{array}{ccc}
 -1 & 0 & 0 \\
 0 & 0 & 1 \\
 0 & -1 & 0 \\
\end{array}
\right) \right\} v_{AC}\,.\nonumber\\
\end{eqnarray}

The matrices in Eq.~\eqref{eq:vacuum_all} are identical to the $\mathbf{3}$-plet representation matrices of the 24 elements of $S_4$ in basis in Table~\ref{tab:rep_matrix_vacuum}. Using the basis transformation, we obtain solutions for $\langle \Phi_{AC} \rangle_{i\alpha} = \sum_{j,\beta} (U_\omega^\dag)_{ij} (U_\omega^\dag)_{\alpha\beta} \langle \tilde{\Phi}_{AC} \rangle_{j\beta}$ in the basis used in the main text, i.e., that in Table~\ref{tab:rep_matrix_main}. With the help of $U_\omega^\dag = U_\omega P_{23}$, we can express these solutions as 
\begin{eqnarray}
\langle \Phi_{AC} \rangle = \rho_{\mathbf{3}}(\gamma) P_{23} v_{AC}
\end{eqnarray} 
with $\gamma$ any element of $S_4$. 

In the main text, we achieved the breaking of modular symmetry $S^A_4\times S^C_4$ to the flavour symmetry $S^D_4$ by assuming the VEV $\langle \Phi_{AC} \rangle_{i\alpha} = (P_{23})_{i\alpha} v_{AC}$ (corresponding to the first solution in Eq.~\eqref{eq:vacuum_all}, $\langle \tilde{\Phi}_{AC} \rangle_{i\alpha} \propto \delta_{i\alpha}$). We comment in the following that this VEV is not special and any VEV with the form $\rho_{\mathbf{3}}(\gamma) P_{23} v_{AC}$ can lead to the breaking of two $S_4$'s to a single $S_4$. 

To see this feature more clearly, let us pick up the second solution in Eq.~\eqref{eq:vacuum_all} as an example. This solution is represented as a $3\times 3$ matrix to be $\langle \Phi_{AC} \rangle = \rho_{\mathbf{3}}(S) P_{23} v_{AC}$, i.e., corresponding to the element $S$. The operator $L \Phi_{AC} Y_A N_A^c H_u$, for instance, after $\Phi_{AC}$ gains the VEV $\rho_{\mathbf{3}}(S) P_{23} v_{AC}$, is effectively expressed as 
\begin{eqnarray}
L \langle \Phi_{AC} \rangle Y_A N_A^c H_u = 
(L_1, L_2, L_3) P_{23} \rho_{\mathbf{3}}(S)
\begin{pmatrix} (Y_A)_1 \\ (Y_A)_2 \\ (Y_A)_3 \end{pmatrix} 
N_A^c H_u v_{AC} \,.
\end{eqnarray}
For any element $\gamma_D$ of $S_4^D$, which leads to $(L_1,L_2,L_3) \to (L_1,L_2,L_3) \rho^T_{\mathbf{3}}(\gamma_D)$, with the help of basis-dependent property $\rho_{\mathbf{3}}^T(\gamma_D) P_{23} = P_{23} \rho_{\mathbf{3}}(\gamma^{-1}_D)$, one can always require the transformation properties for $Y_A$ and $N_A^c$ in the following,
\begin{eqnarray} \label{eq:transformation_rotation}
Y_A &\to& (c_D \tau_A + d_D )^6 \rho_{\mathbf{3}}(S^{-1} \gamma_D S) \, Y_A \,, \nonumber\\
N_A^c &\to& (c_D \tau_A + d_D )^{-6} N_A^c \,,
\end{eqnarray}
such that the effective operator $L \langle \Phi_{AC} \rangle Y_A N_A H_u$ is invariant in $S_4^D$.  
The transformation property for $Y_A$ is equivalent to the following transformation property for $[\rho_{\mathbf{3}}(S) Y_A]$, 
\begin{eqnarray}
[\rho_{\mathbf{3}}(S) Y_A] \to (c_D \tau_A + d_D )^6 \rho_{\mathbf{3}}(\gamma_D) [\rho_{\mathbf{3}}(S) Y_A]\,. 
\end{eqnarray}

Similarly, $S^B_4\times S^C_4$ can be broken to $S^D_4$ by the VEV $\langle \Phi_{BC} \rangle_{m\alpha} = (P_{23})_{m\alpha} v_{AC}$. Eventually, we realise the breaking of three $S_4$'s to a single $S_4$.


\begin{thebibliography}{99}

%\cite{Minkowski:1977sc}
\bibitem{Minkowski:1977sc}
  P.~Minkowski,
  %``$\mu \to e\gamma$ at a Rate of One Out of $10^{9}$ Muon Decays?,''
  Phys.\ Lett.\  {\bf 67B} (1977) 421.
  %doi:10.1016/0370-2693(77)90435-X
  %%CITATION = doi:10.1016/0370-2693(77)90435-X;%%
  %3404 citations counted in INSPIRE as of 04 Jun 2019


	 \bibitem{Yanagida:1979ss}
	 T. Yanagida, In {\it Proceedings of the Workshop on Unified Theory and
	 	the Baryon Number of the Universe}, edited by O. Sawada and A.
	 Sugamoto, (KEK, Tsukuba, 1979), p. 95.
	 
	 \bibitem{Gell-Mann:1979ss}
	 M. Gell-Mann, P. Ramond and R. Slansky, In {\it Supergravity},
	 edited by P. van Nieuwenhuizen and D. Z. Freeman, (North-Holland,
	 Amsterdam, 1979), p. 315.
	 
	 \bibitem{Glashow:1979ss}
	 S. L. Glashow, In {\it Quarks and Leptons}, edited by M. Levy {\it
	 	et al.} (Plenum, New York, 1980), p. 707.


%\cite{Mohapatra:1979ia}
\bibitem{Mohapatra:1979ia} 
  R.~N.~Mohapatra and G.~Senjanovic,
  %``Neutrino Mass and Spontaneous Parity Nonconservation,''
  Phys.\ Rev.\ Lett.\  {\bf 44}, 912 (1980).
  doi:10.1103/PhysRevLett.44.912
  %%CITATION = doi:10.1103/PhysRevLett.44.912;%%
  %5120 citations counted in INSPIRE as of 04 Nov 2019


%\cite{Schechter:1980gr}
\bibitem{Schechter:1980gr} 
  J.~Schechter and J.~W.~F.~Valle,
  %``Neutrino Masses in SU(2) x U(1) Theories,''
  Phys.\ Rev.\ D {\bf 22}, 2227 (1980).
  doi:10.1103/PhysRevD.22.2227
  %%CITATION = doi:10.1103/PhysRevD.22.2227;%%
  %2637 citations counted in INSPIRE as of 04 Nov 2019


%\cite{Schechter:1981cv}
\bibitem{Schechter:1981cv} 
  J.~Schechter and J.~W.~F.~Valle,
  %``Neutrino Decay and Spontaneous Violation of Lepton Number,''
  Phys.\ Rev.\ D {\bf 25}, 774 (1982).
  doi:10.1103/PhysRevD.25.774
  %%CITATION = doi:10.1103/PhysRevD.25.774;%%
  %868 citations counted in INSPIRE as of 04 Nov 2019


%\cite{King:1999mb}
\bibitem{King:1999mb} 
  S.~F.~King,
  %``Large mixing angle MSW and atmospheric neutrinos from single right-handed neutrino dominance and U(1) family symmetry,''
  Nucl.\ Phys.\ B {\bf 576}, 85 (2000)
  doi:10.1016/S0550-3213(00)00109-7
  [hep-ph/9912492].
  %%CITATION = doi:10.1016/S0550-3213(00)00109-7;%%
  %246 citations counted in INSPIRE as of 04 Nov 2019


%\cite{King:2002nf}
\bibitem{King:2002nf} 
  S.~F.~King,
  %``Constructing the large mixing angle MNS matrix in seesaw models with right-handed neutrino dominance,''
  JHEP {\bf 0209}, 011 (2002)
  doi:10.1088/1126-6708/2002/09/011
  [hep-ph/0204360].
  %%CITATION = doi:10.1088/1126-6708/2002/09/011;%%
  %210 citations counted in INSPIRE as of 04 Nov 2019


%\cite{King:2013eh}
\bibitem{King:2013eh} 
  S.~F.~King and C.~Luhn,
  %``Neutrino Mass and Mixing with Discrete Symmetry,''
  Rept.\ Prog.\ Phys.\  {\bf 76}, 056201 (2013)
  doi:10.1088/0034-4885/76/5/056201
  [arXiv:1301.1340 [hep-ph]].
  %%CITATION = doi:10.1088/0034-4885/76/5/056201;%%
  %535 citations counted in INSPIRE as of 04 Nov 2019


%\cite{King:2017guk}
\bibitem{King:2017guk} 
  S.~F.~King,
  %``Unified Models of Neutrinos, Flavour and CP Violation,''
  Prog.\ Part.\ Nucl.\ Phys.\  {\bf 94}, 217 (2017)
  doi:10.1016/j.ppnp.2017.01.003
  [arXiv:1701.04413 [hep-ph]].
  %%CITATION = doi:10.1016/j.ppnp.2017.01.003;%%
  %87 citations counted in INSPIRE as of 04 Nov 2019


%\cite{Varzielas:2012pa}
\bibitem{Varzielas:2012pa} 
  I.~de Medeiros Varzielas and L.~Lavoura,
  %``Flavour models for $TM_{1}$ lepton mixing,''
  J.\ Phys.\ G {\bf 40}, 085002 (2013)
  doi:10.1088/0954-3899/40/8/085002
  [arXiv:1212.3247 [hep-ph]].
  %%CITATION = doi:10.1088/0954-3899/40/8/085002;%%
  %53 citations counted in INSPIRE as of 04 Nov 2019


%\cite{Luhn:2013vna}
\bibitem{Luhn:2013vna} 
  C.~Luhn,
  %``Trimaximal TM$_{1}$ neutrino mixing in S$_{4}$ with spontaneous CP violation,''
  Nucl.\ Phys.\ B {\bf 875}, 80 (2013)
  doi:10.1016/j.nuclphysb.2013.07.003
  [arXiv:1306.2358 [hep-ph]].
  %%CITATION = doi:10.1016/j.nuclphysb.2013.07.003;%%
  %75 citations counted in INSPIRE as of 04 Nov 2019


%\cite{deMedeirosVarzielas:2005qg}
\bibitem{deMedeirosVarzielas:2005qg} 
  I.~de Medeiros Varzielas, S.~F.~King and G.~G.~Ross,
  %``Tri-bimaximal neutrino mixing from discrete subgroups of SU(3) and SO(3) family symmetry,''
  Phys.\ Lett.\ B {\bf 644}, 153 (2007)
  doi:10.1016/j.physletb.2006.11.015
  [hep-ph/0512313].
  %%CITATION = doi:10.1016/j.physletb.2006.11.015;%%
  %208 citations counted in INSPIRE as of 04 Nov 2019


%\cite{Koide:2007sr}
\bibitem{Koide:2007sr} 
  Y.~Koide,
  %``S(4) flavor symmetry embedded into SU(3) and lepton masses and mixing,''
  JHEP {\bf 0708}, 086 (2007)
  doi:10.1088/1126-6708/2007/08/086
  [arXiv:0705.2275 [hep-ph]].
  %%CITATION = doi:10.1088/1126-6708/2007/08/086;%%
  %95 citations counted in INSPIRE as of 04 Nov 2019


%\cite{Banks:2010zn}
\bibitem{Banks:2010zn} 
  T.~Banks and N.~Seiberg,
  %``Symmetries and Strings in Field Theory and Gravity,''
  Phys.\ Rev.\ D {\bf 83}, 084019 (2011)
  doi:10.1103/PhysRevD.83.084019
  [arXiv:1011.5120 [hep-th]].
  %%CITATION = doi:10.1103/PhysRevD.83.084019;%%
  %352 citations counted in INSPIRE as of 04 Nov 2019


%\cite{Luhn:2011ip}
\bibitem{Luhn:2011ip} 
  C.~Luhn,
  %``Spontaneous breaking of SU(3) to finite family symmetries: a pedestrian's approach,''
  JHEP {\bf 1103}, 108 (2011)
  doi:10.1007/JHEP03(2011)108
  [arXiv:1101.2417 [hep-ph]].
  %%CITATION = doi:10.1007/JHEP03(2011)108;%%
  %38 citations counted in INSPIRE as of 04 Nov 2019


%\cite{Merle:2011vy}
\bibitem{Merle:2011vy} 
  A.~Merle and R.~Zwicky,
  %``Explicit and spontaneous breaking of SU(3) into its finite subgroups,''
  JHEP {\bf 1202}, 128 (2012)
  doi:10.1007/JHEP02(2012)128
  [arXiv:1110.4891 [hep-ph]].
  %%CITATION = doi:10.1007/JHEP02(2012)128;%%
  %42 citations counted in INSPIRE as of 04 Nov 2019


%\cite{Wu:2012ria}
\bibitem{Wu:2012ria} 
  Y.~L.~Wu,
  %``SU(3) Gauge Family Symmetry and Prediction for the Lepton-Flavor Mixing and Neutrino Masses with Maximal Spontaneous CP Violation,''
  Phys.\ Lett.\ B {\bf 714}, 286 (2012)
  doi:10.1016/j.physletb.2012.07.020
  [arXiv:1203.2382 [hep-ph]].
  %%CITATION = doi:10.1016/j.physletb.2012.07.020;%%
  %25 citations counted in INSPIRE as of 04 Nov 2019


%\cite{Rachlin:2017rvm}
\bibitem{Rachlin:2017rvm} 
  B.~L.~Rachlin and T.~W.~Kephart,
  %``Spontaneous Breaking of Gauge Groups to Discrete Symmetries,''
  JHEP {\bf 1708}, 110 (2017)
  doi:10.1007/JHEP08(2017)110
  [arXiv:1702.08073 [hep-ph]].
  %%CITATION = doi:10.1007/JHEP08(2017)110;%%
  %8 citations counted in INSPIRE as of 04 Nov 2019


%\cite{King:2018fke}
\bibitem{King:2018fke} 
  S.~F.~King and Y.~L.~Zhou,
  %``Spontaneous breaking of $SO(3)$ to finite family symmetries with supersymmetry - an $A_4$ model,''
  JHEP {\bf 1811}, 173 (2018)
  doi:10.1007/JHEP11(2018)173
  [arXiv:1809.10292 [hep-ph]].
  %%CITATION = doi:10.1007/JHEP11(2018)173;%%
  %13 citations counted in INSPIRE as of 04 Nov 2019


%\cite{Asaka:2001eh}
\bibitem{Asaka:2001eh} 
  T.~Asaka, W.~Buchmuller and L.~Covi,
  %``Gauge unification in six-dimensions,''
  Phys.\ Lett.\ B {\bf 523}, 199 (2001)
  doi:10.1016/S0370-2693(01)01324-7
  [hep-ph/0108021].
  %%CITATION = doi:10.1016/S0370-2693(01)01324-7;%%
  %227 citations counted in INSPIRE as of 04 Nov 2019


%\cite{Altarelli:2006kg}
\bibitem{Altarelli:2006kg} 
  G.~Altarelli, F.~Feruglio and Y.~Lin,
  %``Tri-bimaximal neutrino mixing from orbifolding,''
  Nucl.\ Phys.\ B {\bf 775}, 31 (2007)
  doi:10.1016/j.nuclphysb.2007.03.042
  [hep-ph/0610165].
  %%CITATION = doi:10.1016/j.nuclphysb.2007.03.042;%%
  %214 citations counted in INSPIRE as of 04 Nov 2019


%\cite{Kobayashi:2006wq}
\bibitem{Kobayashi:2006wq} 
  T.~Kobayashi, H.~P.~Nilles, F.~Ploger, S.~Raby and M.~Ratz,
  %``Stringy origin of non-Abelian discrete flavor symmetries,''
  Nucl.\ Phys.\ B {\bf 768}, 135 (2007)
  doi:10.1016/j.nuclphysb.2007.01.018
  [hep-ph/0611020].
  %%CITATION = doi:10.1016/j.nuclphysb.2007.01.018;%%
  %216 citations counted in INSPIRE as of 04 Nov 2019


%\cite{Altarelli:2008bg}
\bibitem{Altarelli:2008bg} 
  G.~Altarelli, F.~Feruglio and C.~Hagedorn,
  %``A SUSY SU(5) Grand Unified Model of Tri-Bimaximal Mixing from A$_4$,''
  JHEP {\bf 0803}, 052 (2008)
  doi:10.1088/1126-6708/2008/03/052
  [arXiv:0802.0090 [hep-ph]].
  %%CITATION = doi:10.1088/1126-6708/2008/03/052;%%
  %183 citations counted in INSPIRE as of 04 Nov 2019


%\cite{Adulpravitchai:2009id}
\bibitem{Adulpravitchai:2009id} 
  A.~Adulpravitchai, A.~Blum and M.~Lindner,
  %``Non-Abelian Discrete Flavor Symmetries from T**2/Z(N) Orbifolds,''
  JHEP {\bf 0907}, 053 (2009)
  doi:10.1088/1126-6708/2009/07/053
  [arXiv:0906.0468 [hep-ph]].
  %%CITATION = doi:10.1088/1126-6708/2009/07/053;%%
  %34 citations counted in INSPIRE as of 04 Nov 2019


%\cite{Burrows:2009pi}
\bibitem{Burrows:2009pi} 
  T.~J.~Burrows and S.~F.~King,
  %``A(4) Family Symmetry from SU(5) SUSY GUTs in 6d,''
  Nucl.\ Phys.\ B {\bf 835}, 174 (2010)
  doi:10.1016/j.nuclphysb.2010.04.002
  [arXiv:0909.1433 [hep-ph]].
  %%CITATION = doi:10.1016/j.nuclphysb.2010.04.002;%%
  %72 citations counted in INSPIRE as of 04 Nov 2019


%\cite{Adulpravitchai:2010na}
\bibitem{Adulpravitchai:2010na} 
  A.~Adulpravitchai and M.~A.~Schmidt,
  %``Flavored Orbifold GUT - an SO(10) x S4 model,''
  JHEP {\bf 1101}, 106 (2011)
  doi:10.1007/JHEP01(2011)106
  [arXiv:1001.3172 [hep-ph]].
  %%CITATION = doi:10.1007/JHEP01(2011)106;%%
  %26 citations counted in INSPIRE as of 04 Nov 2019


%\cite{Burrows:2010wz}
\bibitem{Burrows:2010wz} 
  T.~J.~Burrows and S.~F.~King,
  %``$A_4$ x SU(5) SUSY GUT of Flavour in 8d,''
  Nucl.\ Phys.\ B {\bf 842}, 107 (2011)
  doi:10.1016/j.nuclphysb.2010.08.018
  [arXiv:1007.2310 [hep-ph]].
  %%CITATION = doi:10.1016/j.nuclphysb.2010.08.018;%%
  %50 citations counted in INSPIRE as of 04 Nov 2019


%\cite{deAnda:2018oik}
\bibitem{deAnda:2018oik} 
  F.~J.~de Anda and S.~F.~King,
  %``An $S_4 \times SU(5)$ SUSY GUT of flavour in 6d,''
  JHEP {\bf 1807}, 057 (2018)
  doi:10.1007/JHEP07(2018)057
  [arXiv:1803.04978 [hep-ph]].
  %%CITATION = doi:10.1007/JHEP07(2018)057;%%
  %21 citations counted in INSPIRE as of 04 Nov 2019


%\cite{Kobayashi:2018rad}
\bibitem{Kobayashi:2018rad} 
  T.~Kobayashi, S.~Nagamoto, S.~Takada, S.~Tamba and T.~H.~Tatsuishi,
  %``Modular symmetry and non-Abelian discrete flavor symmetries in string compactification,''
  Phys.\ Rev.\ D {\bf 97}, no. 11, 116002 (2018)
  doi:10.1103/PhysRevD.97.116002
  [arXiv:1804.06644 [hep-th]].
  %%CITATION = doi:10.1103/PhysRevD.97.116002;%%
  %26 citations counted in INSPIRE as of 04 Nov 2019


%\cite{deAnda:2018yfp}
\bibitem{deAnda:2018yfp} 
  F.~J.~de Anda and S.~F.~King,
  %``$SU(3) \times SO(10)$ in 6d,''
  JHEP {\bf 1810}, 128 (2018)
  doi:10.1007/JHEP10(2018)128
  [arXiv:1807.07078 [hep-ph]].
  %%CITATION = doi:10.1007/JHEP10(2018)128;%%
  %10 citations counted in INSPIRE as of 04 Nov 2019


%\cite{Baur:2019kwi}
\bibitem{Baur:2019kwi} 
  A.~Baur, H.~P.~Nilles, A.~Trautner and P.~K.~S.~Vaudrevange,
  %``Unification of Flavor, CP, and Modular Symmetries,''
  Phys.\ Lett.\ B {\bf 795}, 7 (2019)
  doi:10.1016/j.physletb.2019.03.066
  [arXiv:1901.03251 [hep-th]].
  %%CITATION = doi:10.1016/j.physletb.2019.03.066;%%
  %26 citations counted in INSPIRE as of 04 Nov 2019


%\cite{Kobayashi:2008ih}
\bibitem{Kobayashi:2008ih} 
  T.~Kobayashi, Y.~Omura and K.~Yoshioka,
  %``Flavor Symmetry Breaking and Vacuum Alignment on Orbifolds,''
  Phys.\ Rev.\ D {\bf 78}, 115006 (2008)
  doi:10.1103/PhysRevD.78.115006
  [arXiv:0809.3064 [hep-ph]].
  %%CITATION = doi:10.1103/PhysRevD.78.115006;%%
  %50 citations counted in INSPIRE as of 04 Nov 2019


%\cite{Olguin-Trejo:2018wpw}
\bibitem{Olguin-Trejo:2018wpw} 
  Y.~Olguin-Trejo, R.~Pérez-Martínez and S.~Ramos-Sánchez,
  %``Charting the flavor landscape of MSSM-like Abelian heterotic orbifolds,''
  Phys.\ Rev.\ D {\bf 98}, no. 10, 106020 (2018)
  doi:10.1103/PhysRevD.98.106020
  [arXiv:1808.06622 [hep-th]].
  %%CITATION = doi:10.1103/PhysRevD.98.106020;%%
  %13 citations counted in INSPIRE as of 04 Nov 2019


%\cite{Mutter:2018sra}
\bibitem{Mutter:2018sra} 
  A.~Mütter, E.~Parr and P.~K.~S.~Vaudrevange,
  %``Deep learning in the heterotic orbifold landscape,''
  Nucl.\ Phys.\ B {\bf 940}, 113 (2019)
  doi:10.1016/j.nuclphysb.2019.01.013
  [arXiv:1811.05993 [hep-th]].
  %%CITATION = doi:10.1016/j.nuclphysb.2019.01.013;%%
  %11 citations counted in INSPIRE as of 04 Nov 2019


%\cite{Giveon:1988tt}
\bibitem{Giveon:1988tt} 
  A.~Giveon, E.~Rabinovici and G.~Veneziano,
  %``Duality in String Background Space,''
  Nucl.\ Phys.\ B {\bf 322}, 167 (1989).
  doi:10.1016/0550-3213(89)90489-6
  %%CITATION = doi:10.1016/0550-3213(89)90489-6;%%
  %345 citations counted in INSPIRE as of 04 Nov 2019


%\cite{Ferrara:1989bc}
\bibitem{Ferrara:1989bc} 
  S.~Ferrara, D.~Lust, A.~D.~Shapere and S.~Theisen,
  %``Modular Invariance in Supersymmetric Field Theories,''
  Phys.\ Lett.\ B {\bf 225}, 363 (1989).
  doi:10.1016/0370-2693(89)90583-2
  %%CITATION = doi:10.1016/0370-2693(89)90583-2;%%
  %233 citations counted in INSPIRE as of 04 Nov 2019


%\cite{Ferrara:1989qb}
\bibitem{Ferrara:1989qb} 
  S.~Ferrara, .D.~Lust and S.~Theisen,
  %``Target Space Modular Invariance and Low-Energy Couplings in Orbifold Compactifications,''
  Phys.\ Lett.\ B {\bf 233}, 147 (1989).
  doi:10.1016/0370-2693(89)90631-X
  %%CITATION = doi:10.1016/0370-2693(89)90631-X;%%
  %192 citations counted in INSPIRE as of 04 Nov 2019


%\cite{deAnda:2018ecu}
\bibitem{deAnda:2018ecu}
  F.~J.~de Anda, S.~F.~King and E.~Perdomo,
  %``$SU(5)$ grand unified theory with $A_4$ modular symmetry,''
  Phys.\ Rev.\ D {\bf 101} (2020) no.1,  015028
  doi:10.1103/PhysRevD.101.015028
  [arXiv:1812.05620 [hep-ph]].
  %%CITATION = doi:10.1103/PhysRevD.101.015028;%%
  %37 citations counted in INSPIRE as of 09 Mar 2020


%\cite{Kobayashi:2018bff}
\bibitem{Kobayashi:2018bff} 
  T.~Kobayashi and S.~Tamba,
  %``Modular forms of finite modular subgroups from magnetized D-brane models,''
  Phys.\ Rev.\ D {\bf 99}, no. 4, 046001 (2019)
  doi:10.1103/PhysRevD.99.046001
  [arXiv:1811.11384 [hep-th]].
  %%CITATION = doi:10.1103/PhysRevD.99.046001;%%
  %20 citations counted in INSPIRE as of 04 Nov 2019


%\cite{Altarelli:2005yx}
\bibitem{Altarelli:2005yx} 
  G.~Altarelli and F.~Feruglio,
  %``Tri-bimaximal neutrino mixing, A(4) and the modular symmetry,''
  Nucl.\ Phys.\ B {\bf 741}, 215 (2006)
  doi:10.1016/j.nuclphysb.2006.02.015
  [hep-ph/0512103].
  %%CITATION = doi:10.1016/j.nuclphysb.2006.02.015;%%
  %600 citations counted in INSPIRE as of 04 Nov 2019


%\cite{deAdelhartToorop:2011re}
\bibitem{deAdelhartToorop:2011re} 
  R.~de Adelhart Toorop, F.~Feruglio and C.~Hagedorn,
  %``Finite Modular Groups and Lepton Mixing,''
  Nucl.\ Phys.\ B {\bf 858}, 437 (2012)
  doi:10.1016/j.nuclphysb.2012.01.017
  [arXiv:1112.1340 [hep-ph]].
  %%CITATION = doi:10.1016/j.nuclphysb.2012.01.017;%%
  %162 citations counted in INSPIRE as of 04 Nov 2019


%\cite{Feruglio:2017spp}
\bibitem{Feruglio:2017spp} 
  F.~Feruglio,
  %``Are neutrino masses modular forms?,''
  doi:10.1142/9789813238053\_0012
  arXiv:1706.08749 [hep-ph].
  %%CITATION = doi:10.1142/9789813238053_0012;%%
  %46 citations counted in INSPIRE as of 04 Nov 2019


%\cite{Criado:2018thu}
\bibitem{Criado:2018thu} 
  J.~C.~Criado and F.~Feruglio,
  %``Modular Invariance Faces Precision Neutrino Data,''
  SciPost Phys.\  {\bf 5}, no. 5, 042 (2018)
  doi:10.21468/SciPostPhys.5.5.042
  [arXiv:1807.01125 [hep-ph]].
  %%CITATION = doi:10.21468/SciPostPhys.5.5.042;%%
  %41 citations counted in INSPIRE as of 04 Nov 2019


%\cite{Kobayashi:2018vbk}
\bibitem{Kobayashi:2018vbk} 
  T.~Kobayashi, K.~Tanaka and T.~H.~Tatsuishi,
  %``Neutrino mixing from finite modular groups,''
  Phys.\ Rev.\ D {\bf 98}, no. 1, 016004 (2018)
  doi:10.1103/PhysRevD.98.016004
  [arXiv:1803.10391 [hep-ph]].
  %%CITATION = doi:10.1103/PhysRevD.98.016004;%%
  %41 citations counted in INSPIRE as of 04 Nov 2019


%\cite{Kobayashi:2018wkl}
\bibitem{Kobayashi:2018wkl} 
  T.~Kobayashi, Y.~Shimizu, K.~Takagi, M.~Tanimoto, T.~H.~Tatsuishi and H.~Uchida,
  %``Finite modular subgroups for fermion mass matrices and baryon/lepton number violation,''
  Phys.\ Lett.\ B {\bf 794}, 114 (2019)
  doi:10.1016/j.physletb.2019.05.034
  [arXiv:1812.11072 [hep-ph]].
  %%CITATION = doi:10.1016/j.physletb.2019.05.034;%%
  %30 citations counted in INSPIRE as of 04 Nov 2019


%\cite{Kobayashi:2018scp}
\bibitem{Kobayashi:2018scp} 
  T.~Kobayashi, N.~Omoto, Y.~Shimizu, K.~Takagi, M.~Tanimoto and T.~H.~Tatsuishi,
  %``Modular A$_{4}$ invariance and neutrino mixing,''
  JHEP {\bf 1811}, 196 (2018)
  doi:10.1007/JHEP11(2018)196
  [arXiv:1808.03012 [hep-ph]].
  %%CITATION = doi:10.1007/JHEP11(2018)196;%%
  %38 citations counted in INSPIRE as of 04 Nov 2019


%\cite{Okada:2018yrn}
\bibitem{Okada:2018yrn} 
  H.~Okada and M.~Tanimoto,
  %``CP violation of quarks in $A_4$ modular invariance,''
  Phys.\ Lett.\ B {\bf 791}, 54 (2019)
  doi:10.1016/j.physletb.2019.02.028
  [arXiv:1812.09677 [hep-ph]].
  %%CITATION = doi:10.1016/j.physletb.2019.02.028;%%
  %27 citations counted in INSPIRE as of 04 Nov 2019


%\cite{Novichkov:2018yse}
\bibitem{Novichkov:2018yse} 
  P.~P.~Novichkov, S.~T.~Petcov and M.~Tanimoto,
  %``Trimaximal Neutrino Mixing from Modular A4 Invariance with Residual Symmetries,''
  Phys.\ Lett.\ B {\bf 793}, 247 (2019)
  doi:10.1016/j.physletb.2019.04.043
  [arXiv:1812.11289 [hep-ph]].
  %%CITATION = doi:10.1016/j.physletb.2019.04.043;%%
  %29 citations counted in INSPIRE as of 04 Nov 2019


%\cite{Penedo:2018nmg}
\bibitem{Penedo:2018nmg} 
  J.~T.~Penedo and S.~T.~Petcov,
  %``Lepton Masses and Mixing from Modular $S_4$ Symmetry,''
  Nucl.\ Phys.\ B {\bf 939}, 292 (2019)
  doi:10.1016/j.nuclphysb.2018.12.016
  [arXiv:1806.11040 [hep-ph]].
  %%CITATION = doi:10.1016/j.nuclphysb.2018.12.016;%%
  %39 citations counted in INSPIRE as of 04 Nov 2019


%\cite{Novichkov:2018ovf}
\bibitem{Novichkov:2018ovf} 
  P.~P.~Novichkov, J.~T.~Penedo, S.~T.~Petcov and A.~V.~Titov,
  %``Modular S$_{4}$ models of lepton masses and mixing,''
  JHEP {\bf 1904}, 005 (2019)
  doi:10.1007/JHEP04(2019)005
  [arXiv:1811.04933 [hep-ph]].
  %%CITATION = doi:10.1007/JHEP04(2019)005;%%
  %34 citations counted in INSPIRE as of 04 Nov 2019


%\cite{Novichkov:2018nkm}
\bibitem{Novichkov:2018nkm} 
  P.~P.~Novichkov, J.~T.~Penedo, S.~T.~Petcov and A.~V.~Titov,
  %``Modular A$_{5}$ symmetry for flavour model building,''
  JHEP {\bf 1904}, 174 (2019)
  doi:10.1007/JHEP04(2019)174
  [arXiv:1812.02158 [hep-ph]].
  %%CITATION = doi:10.1007/JHEP04(2019)174;%%
  %31 citations counted in INSPIRE as of 04 Nov 2019


%\cite{Ding:2019xna}
\bibitem{Ding:2019xna}
  G.~J.~Ding, S.~F.~King and X.~G.~Liu,
  %``Neutrino mass and mixing with $A_5$ modular symmetry,''
  Phys.\ Rev.\ D {\bf 100} (2019) no.11,  115005
  doi:10.1103/PhysRevD.100.115005
  [arXiv:1903.12588 [hep-ph]].
  %%CITATION = doi:10.1103/PhysRevD.100.115005;%%
  %35 citations counted in INSPIRE as of 09 Mar 2020


%\cite{Okada:2019uoy}
\bibitem{Okada:2019uoy} 
  H.~Okada and M.~Tanimoto,
  %``Towards unification of quark and lepton flavors in $A_4$ modular invariance,''
  arXiv:1905.13421 [hep-ph].
  %%CITATION = ARXIV:1905.13421;%%
  %18 citations counted in INSPIRE as of 04 Nov 2019


%\cite{Ding:2018fyz}
\bibitem{Ding:2018fyz} 
  G.~J.~Ding, S.~F.~King and C.~C.~Li,
  %``Tri-Direct CP in the Littlest Seesaw Playground,''
  JHEP {\bf 1812}, 003 (2018)
  doi:10.1007/JHEP12(2018)003
  [arXiv:1807.07538 [hep-ph]].
  %%CITATION = doi:10.1007/JHEP12(2018)003;%%
  %10 citations counted in INSPIRE as of 04 Nov 2019


%\cite{Ding:2018tuj}
\bibitem{Ding:2018tuj} 
  G.~J.~Ding, S.~F.~King and C.~C.~Li,
  %``Lepton mixing predictions from $S_4$ in the tridirect CP approach to two right-handed neutrino models,''
  Phys.\ Rev.\ D {\bf 99}, no. 7, 075035 (2019)
  doi:10.1103/PhysRevD.99.075035
  [arXiv:1811.12340 [hep-ph]].
  %%CITATION = doi:10.1103/PhysRevD.99.075035;%%
  %7 citations counted in INSPIRE as of 04 Nov 2019


%\cite{Novichkov:2019sqv}
\bibitem{Novichkov:2019sqv} 
  P.~P.~Novichkov, J.~T.~Penedo, S.~T.~Petcov and A.~V.~Titov,
  %``Generalised CP Symmetry in Modular-Invariant Models of Flavour,''
  JHEP {\bf 1907}, 165 (2019)
  doi:10.1007/JHEP07(2019)165
  [arXiv:1905.11970 [hep-ph]].
  %%CITATION = doi:10.1007/JHEP07(2019)165;%%
  %18 citations counted in INSPIRE as of 04 Nov 2019


%\cite{Ibanez:1986ka}
\bibitem{Ibanez:1986ka} 
  L.~E.~Ibanez,
  %``Hierarchy of Quark - Lepton Masses in Orbifold Superstring Compactification,''
  Phys.\ Lett.\ B {\bf 181}, 269 (1986).
  doi:10.1016/0370-2693(86)90044-4
  %%CITATION = doi:10.1016/0370-2693(86)90044-4;%%
  %74 citations counted in INSPIRE as of 04 Nov 2019


%\cite{Casas:1991ac}
\bibitem{Casas:1991ac} 
  J.~A.~Casas, F.~Gomez and C.~Munoz,
  %``Complete structure of Z(n) Yukawa couplings,''
  Int.\ J.\ Mod.\ Phys.\ A {\bf 8}, 455 (1993)
  doi:10.1142/S0217751X93000187
  [hep-th/9110060].
  %%CITATION = doi:10.1142/S0217751X93000187;%%
  %66 citations counted in INSPIRE as of 04 Nov 2019


%\cite{Lebedev:2001qg}
\bibitem{Lebedev:2001qg} 
  O.~Lebedev,
  %``The CKM phase in heterotic orbifold models,''
  Phys.\ Lett.\ B {\bf 521}, 71 (2001)
  doi:10.1016/S0370-2693(01)01180-7
  [hep-th/0108218].
  %%CITATION = doi:10.1016/S0370-2693(01)01180-7;%%
  %26 citations counted in INSPIRE as of 04 Nov 2019


%\cite{Kobayashi:2003vi}
\bibitem{Kobayashi:2003vi} 
  T.~Kobayashi and O.~Lebedev,
  %``Heterotic Yukawa couplings and continuous Wilson lines,''
  Phys.\ Lett.\ B {\bf 566}, 164 (2003)
  doi:10.1016/S0370-2693(03)00560-4
  [hep-th/0303009].
  %%CITATION = doi:10.1016/S0370-2693(03)00560-4;%%
  %24 citations counted in INSPIRE as of 04 Nov 2019


%\cite{Brax:1994kv}
\bibitem{Brax:1994kv} 
  P.~Brax and M.~Chemtob,
  %``Flavor changing neutral current constraints on standard - like orbifold models,''
  Phys.\ Rev.\ D {\bf 51}, 6550 (1995)
  doi:10.1103/PhysRevD.51.6550
  [hep-th/9411022].
  %%CITATION = doi:10.1103/PhysRevD.51.6550;%%
  %25 citations counted in INSPIRE as of 04 Nov 2019


%\cite{Binetruy:1995nt}
\bibitem{Binetruy:1995nt} 
  P.~Binetruy and E.~Dudas,
  %``Dynamical mass matrices from effective superstring theories,''
  Nucl.\ Phys.\ B {\bf 451}, 31 (1995)
  doi:10.1016/0550-3213(95)00345-S
  [hep-ph/9505295].
  %%CITATION = doi:10.1016/0550-3213(95)00345-S;%%
  %21 citations counted in INSPIRE as of 04 Nov 2019


%\cite{Dudas:1995eq}
\bibitem{Dudas:1995eq} 
  E.~Dudas, S.~Pokorski and C.~A.~Savoy,
  %``Soft scalar masses in supergravity with horizontal U(1)-x gauge symmetry,''
  Phys.\ Lett.\ B {\bf 369}, 255 (1996)
  doi:10.1016/0370-2693(95)01536-1
  [hep-ph/9509410].
  %%CITATION = doi:10.1016/0370-2693(95)01536-1;%%
  %136 citations counted in INSPIRE as of 04 Nov 2019


%\cite{Dudas:1996aa}
\bibitem{Dudas:1996aa} 
  E.~Dudas,
  %``Dynamical mass matrices from moduli fields,''
  hep-ph/9602231.
  %%CITATION = HEP-PH/9602231;%%
  %5 citations counted in INSPIRE as of 04 Nov 2019


%\cite{Leontaris:1997vw}
\bibitem{Leontaris:1997vw} 
  G.~K.~Leontaris and N.~D.~Tracas,
  %``Modular weights, U(1)'s and mass matrices,''
  Phys.\ Lett.\ B {\bf 419}, 206 (1998)
  doi:10.1016/S0370-2693(97)01412-3
  [hep-ph/9709510].
  %%CITATION = doi:10.1016/S0370-2693(97)01412-3;%%
  %17 citations counted in INSPIRE as of 04 Nov 2019


%\cite{Dent:2001cc}
\bibitem{Dent:2001cc} 
  T.~Dent,
  %``CP violation and modular symmetries,''
  Phys.\ Rev.\ D {\bf 64}, 056005 (2001)
  doi:10.1103/PhysRevD.64.056005
  [hep-ph/0105285].
  %%CITATION = doi:10.1103/PhysRevD.64.056005;%%
  %18 citations counted in INSPIRE as of 04 Nov 2019


%\cite{Dent:2001mn}
\bibitem{Dent:2001mn} 
  T.~Dent,
  %``On the modular invariance of mass eigenstates and CP violation,''
  JHEP {\bf 0112}, 028 (2001)
  doi:10.1088/1126-6708/2001/12/028
  [hep-th/0111024].
  %%CITATION = doi:10.1088/1126-6708/2001/12/028;%%
  %7 citations counted in INSPIRE as of 04 Nov 2019


%\cite{King:2004tx}
\bibitem{King:2004tx} 
  S.~F.~King, I.~N.~R.~Peddie, G.~G.~Ross, L.~Velasco-Sevilla and O.~Vives,
  %``Kahler corrections and softly broken family symmetries,''
  JHEP {\bf 0507}, 049 (2005)
  doi:10.1088/1126-6708/2005/07/049
  [hep-ph/0407012].
  %%CITATION = doi:10.1088/1126-6708/2005/07/049;%%
  %44 citations counted in INSPIRE as of 04 Nov 2019


%\cite{Balasubramanian:2004uy}
\bibitem{Balasubramanian:2004uy} 
  V.~Balasubramanian and P.~Berglund,
  %``Stringy corrections to Kahler potentials, SUSY breaking, and the cosmological constant problem,''
  JHEP {\bf 0411}, 085 (2004)
  doi:10.1088/1126-6708/2004/11/085
  [hep-th/0408054].
  %%CITATION = doi:10.1088/1126-6708/2004/11/085;%%
  %180 citations counted in INSPIRE as of 04 Nov 2019


%\cite{Silverstein:2004id}
\bibitem{Silverstein:2004id} 
  E.~Silverstein,
  %``TASI / PiTP / ISS lectures on moduli and microphysics,''
  doi:10.1142/9789812775108\_0004
  hep-th/0405068.
  %%CITATION = doi:10.1142/9789812775108_0004;%%
  %107 citations counted in INSPIRE as of 04 Nov 2019


%\cite{Xing:2006ms}
\bibitem{Xing:2006ms} 
  Z.~z.~Xing and S.~Zhou,
  %``Tri-bimaximal Neutrino Mixing and Flavor-dependent Resonant Leptogenesis,''
  Phys.\ Lett.\ B {\bf 653}, 278 (2007)
  doi:10.1016/j.physletb.2007.08.009
  [hep-ph/0607302].
  %%CITATION = doi:10.1016/j.physletb.2007.08.009;%%
  %80 citations counted in INSPIRE as of 04 Nov 2019


%\cite{Lam:2006wm}
\bibitem{Lam:2006wm} 
  C.~S.~Lam,
  %``Mass Independent Textures and Symmetry,''
  Phys.\ Rev.\ D {\bf 74}, 113004 (2006)
  doi:10.1103/PhysRevD.74.113004
  [hep-ph/0611017].
  %%CITATION = doi:10.1103/PhysRevD.74.113004;%%
  %74 citations counted in INSPIRE as of 04 Nov 2019


%\cite{Albright:2008rp}
\bibitem{Albright:2008rp} 
  C.~H.~Albright and W.~Rodejohann,
  %``Comparing Trimaximal Mixing and Its Variants with Deviations from Tri-bimaximal Mixing,''
  Eur.\ Phys.\ J.\ C {\bf 62}, 599 (2009)
  doi:10.1140/epjc/s10052-009-1074-3
  [arXiv:0812.0436 [hep-ph]].
  %%CITATION = doi:10.1140/epjc/s10052-009-1074-3;%%
  %120 citations counted in INSPIRE as of 04 Nov 2019


%\cite{Albright:2010ap}
\bibitem{Albright:2010ap} 
  C.~H.~Albright, A.~Dueck and W.~Rodejohann,
  %``Possible Alternatives to Tri-bimaximal Mixing,''
  Eur.\ Phys.\ J.\ C {\bf 70}, 1099 (2010)
  doi:10.1140/epjc/s10052-010-1492-2
  [arXiv:1004.2798 [hep-ph]].
  %%CITATION = doi:10.1140/epjc/s10052-010-1492-2;%%
  %175 citations counted in INSPIRE as of 04 Nov 2019


%\cite{King:2015dvf}
\bibitem{King:2015dvf} 
  S.~F.~King,
  %``Littlest Seesaw,''
  JHEP {\bf 1602}, 085 (2016)
  doi:10.1007/JHEP02(2016)085
  [arXiv:1512.07531 [hep-ph]].
  %%CITATION = doi:10.1007/JHEP02(2016)085;%%
  %39 citations counted in INSPIRE as of 04 Nov 2019


%\cite{Esteban:2018azc}
\bibitem{Esteban:2018azc} 
  I.~Esteban, M.~C.~Gonzalez-Garcia, A.~Hernandez-Cabezudo, M.~Maltoni and T.~Schwetz,
  %``Global analysis of three-flavour neutrino oscillations: synergies and tensions in the determination of $\theta_{23}$, $\delta_{CP}$, and the mass ordering,''
  JHEP {\bf 1901}, 106 (2019)
  doi:10.1007/JHEP01(2019)106
  [arXiv:1811.05487 [hep-ph]].
  %%CITATION = doi:10.1007/JHEP01(2019)106;%%
  %201 citations counted in INSPIRE as of 04 Nov 2019


\bibitem{nufit4}
NuFIT 4.0 (2018), www.nu-fit.org.

%\cite{Hirsch:2010ru}
\bibitem{Hirsch:2010ru} 
  M.~Hirsch, S.~Morisi, E.~Peinado and J.~W.~F.~Valle,
  %``Discrete dark matter,''
  Phys.\ Rev.\ D {\bf 82}, 116003 (2010)
  doi:10.1103/PhysRevD.82.116003
  [arXiv:1007.0871 [hep-ph]].
  %%CITATION = doi:10.1103/PhysRevD.82.116003;%%
  %100 citations counted in INSPIRE as of 09 Mar 2020

%\cite{Escobar:2008vc}
\bibitem{Escobar:2008vc}
  J.~A.~Escobar and C.~Luhn,
  %``The Flavor Group Delta(6n**2),''
  J.\ Math.\ Phys.\  {\bf 50} (2009) 013524
  %doi:10.1063/1.3046563
  [arXiv:0809.0639 [hep-th]].
  %%CITATION = doi:10.1063/1.3046563;%%
  %96 citations counted in INSPIRE as of 04 Jun 2019
 
\end{thebibliography}
\end{document}